%% file: sample-sigconf.tex
\makeatletter
\newcommand{\linebreakand}{%
  \end{@IEEEauthorhalign}
  \hfill\mbox{}\par
  \mbox{}\hfill\begin{@IEEEauthorhalign}
}
\makeatother

\documentclass[conference]{IEEEtran}
\usepackage{amsmath,amsfonts}
\usepackage{algorithmic}
\usepackage{algorithm}
\usepackage{array}
\usepackage{textcomp}
\usepackage{stfloats}
\usepackage{url}
\usepackage{verbatim}
\usepackage{graphicx}
\usepackage{cite}
\AtBeginDocument{%
  \providecommand\BibTeX{{%
    Bib\TeX}}}

\usepackage{tikz}
\usepackage{amsmath}

\usepackage{graphicx}
\usepackage{subcaption}

\usepackage{filecontents}
\usepackage{adjustbox}

\usepackage{hhline}
\usepackage{array}

\usepackage{tikz}
\usepackage{amsmath}
\usepackage{multirow}
\usepackage{xurl}
\usepackage{hyperref}
\hypersetup{
    breaklinks=true,
    colorlinks=false,   
    hidelinks          
}

\usepackage{caption}

\newcommand{\szz}{D-SZZ}

\usepackage{filecontents}

\usepackage{listings}
\usepackage[dvipsnames]{xcolor}
\usepackage{enumitem}

\def\BibTeX{{\rm B\kern-.05em{\sc i\kern-.025em b}\kern-.08em
    T\kern-.1667em\lower.7ex\hbox{E}\kern-.125emX}}

\begin{document}

\title{Exploring Early Vulnerability Analysis via Automatic Identification of Vulnerability-Introducing Commits}


% % \thanks{This paper was produced by the IEEE Publication Technology Group. They are in Piscataway, NJ.}% <-this % stops a space
% % \thanks{Manuscript received April 19, 2021; revised August 16, 2021.}}

\author{
\IEEEauthorblockN{Weiliang Qi}
\IEEEauthorblockA{\textit{The University of Texas at Dallas}\\
Weiliang.Qi@utdallas.edu}
\and
\IEEEauthorblockN{Shu Wang}
\IEEEauthorblockA{\textit{Palo Alto Networks}\\
shuwang@paloaltonetworks.com}
\and
\IEEEauthorblockN{Youpeng Li}
\IEEEauthorblockA{\textit{The University of Texas at Dallas}\\
Youpeng.Li@UTDallas.edu}
\linebreakand
\IEEEauthorblockN{Ela Guo}
\IEEEauthorblockA{\textit{The University of Texas at Dallas}\\
Ela.Guo@UTDallas.edu}
\and
\IEEEauthorblockN{Xinda Wang}
\IEEEauthorblockA{\textit{The University of Texas at Dallas}\\
Xinda.Wang@UTDallas.edu}
}


\newcommand{\sw}[1]{\textcolor{red}{\{\textbf{SW:} #1\}}}

\newcommand{\swsolved}[1]{\textcolor{olive}{\{\textbf{SW:} #1\}}}

\newcommand{\wl}[1]{\textcolor{orange}{\{\textbf{WL:} #1\}}}
\renewcommand{\figureautorefname}{Fig.}



\maketitle
\input{sections/0abstract}


\input{sections/1introduction}
\input{sections/2preliminaries}

\input{sections/3motivation}

\input{sections/3x5Methodology}

\input{sections/4design}

\input{sections/5evaluation}

\input{sections/8discussion}

\input{sections/9conclusion}


\bibliographystyle{IEEEtran}
\bibliography{sample-base}


\input{sections/10dataavailibility}

\appendix
\input{sections/appendix}

\end{document}

%% file: sections/0abstract.tex
\begin{abstract}

Developing effective vulnerability analysis techniques requires high-quality datasets that capture how vulnerabilities appear and evolve in real software codebase. Existing datasets, however, typically label as vulnerable only the code immediately preceding vulnerability-fixing commits (VFCs), overlooking the earlier versions in which vulnerabilities were first introduced and subsequently evolved. This practice limits both dataset coverage and our understanding of whether current detectors can identify vulnerabilities during their early, latent stages. In this paper, we present three manually annotated datasets of vulnerability-introducing commits (VICs) based on recent CVEs and widely used open-source projects. We further propose D-SZZ, a data state-aware VIC identification framework that analyzes security-fixing patterns and performs vulnerability scope slicing to recover the semantics implied by a fix. Experimental results demonstrate that D-SZZ outperforms existing SZZ and LLM-based baselines in identifying VICs. In addition, we evaluate six representative vulnerability detectors and find that models trained only on pre-VFC code struggle with original vulnerabilities, while incorporating early-stage samples improves detection performance. We also show that D-SZZ improves downstream vulnerable-version identification, and use it to uncover 16 inaccurate affected-version records in the NVD, all of which were officially confirmed and updated.

\end{abstract}

%% file: sections/1introduction.tex
\section{Introduction}
Software vulnerabilities continue to proliferate at an unprecedented rate, posing significant risks to modern computing systems~\cite{durumeric2014matter, luszcz2018apache, hiesgen2022race}. Once exploited, these vulnerabilities can lead to unauthorized access, data breaches, system disruptions, and compromise of critical systems. To mitigate these threats, researchers have developed a wide range of vulnerability analysis techniques~\cite{wu2022vulcnn, fu2022linevul, zhou2019devign, chakraborty2021deep}. However, developing and benchmarking such techniques depends on the availability of large-scale, high-quality vulnerability datasets.

Despite extensive efforts to construct real-world vulnerability datasets from open-source software (OSS)~\cite{fan2020ac, chen2023diversevul, ding2024vulnerability, wang2021patchdb}, a critical limitation persists: existing datasets typically label only the code immediately preceding security-fixing commits (VFCs) as vulnerable. This practice overlooks the historical evolution of vulnerabilities, discarding numerous intermediate vulnerable versions between vulnerability introduction and remediation. Omitting these latent vulnerability samples not only limits dataset scale but also fails to capture code patterns that emerge during the early dormant period of vulnerabilities. Consequently, models trained on such narrowly scoped data may struggle to detect early-stage vulnerabilities in a timely manner due to feature distribution shifts, leaving systems exposed to attackers for prolonged periods. To collect such early-stage samples, it is critical to pinpoint the precise origins of vulnerabilities, i.e., vulnerability-introducing commits (VICs).

Prior work~\cite{fan2019impact} has leveraged the SZZ algorithm~\cite{sliwerski2005changes} to identify bug-introducing commits, thereby determining VICs and collecting the original vulnerable code.
SZZ operates on the assumption that all lines modified in a fixing commit are essential to the patch, tracing them back towards their origin to pinpoint the defect's introduction. However, real-world fixing commits are frequently entangled, containing modifications unrelated to the security remediation~\cite{davies2014comparing, kim2006automatic} ranging from whitespace adjustment and code formatting to comment updates, which severely reduces its accuracy. To address this, subsequent research have enhanced SZZ by filtering out meta-changes~\cite{da2016framework}, analyzing program dependencies~\cite{tang2024enhancing}, or applying deep learning~\cite{tang2023neural,bao2022vszz} and large language models (LLMs)~\cite{tang2025llm4szz} to identify the defect-related statements. 

However, existing VIC identification methods have two major limitations. %, \emph{limited search scope} and \emph{domain mismatch}.
First, while SZZ has been utilized to identify vulnerability-introducing commits, existing methods such as V-SZZ~\cite{bao2022vszz} restrict their search scope exclusively to deleted lines within fixing commits.
However, vulnerabilities stem not only from defective existing code but also from the absence of critical logic.
Therefore, it cannot address scenarios where remediation only involves code additions (e.g., inserting security checks), inherently limiting the model recall.
Second, existing SZZ-based approaches~\cite{tang2023neural, tang2024enhancing, tang2025llm4szz} are primarily calibrated to detect general buggy statements. 
Since vulnerabilities are a special class of bugs with distinct syntactic and semantic characteristics~\cite{camilo2015bugs}, applying these general-purpose methods to security contexts can lead to suboptimal precision due to domain mismatch.

To solve these challenges, we propose \szz{}, a novel framework that detects VICs by analyzing semantics of VFCs and tracing their software repository history. 
To eliminate the noise inherent in entangled commits, \szz{} first constructs Code Property Graph (CPG) of the VFC and leverages security-fixing patterns to strictly isolate statements related to the vulnerability fix.
Then, instead of tracking only deleted lines, we use data- and control-flow analysis to identify the pre-fix data states that causally interact with these statements and capture them as vulnerability semantics. This semantic-driven approach enables \szz{} to handle fix-by-addition scenarios by tracing the logic implied by the fix rather than the explicitly removed lines.
Finally, \szz{} backtracks these semantics through the repository history until locating the VIC.

To enable comprehensive evaluation, we constructed the three manually annotated VIC datasets, comprising 783 real-world C/C++ samples curated from recent National Vulnerability Database (NVD) records and two widely deployed OSS projects (i.e., tcpdump and Vim).
Experimental results demonstrate that \szz{} significantly outperforms the best baseline, V-SZZ, with recall improvements of 6.63\%--13.64\%. We also observe similar improvements when comparing D-SZZ with other SZZ methods on an external dataset collected from Linux used in previous works. Further, we evaluate \szz{} on the downstream task of vulnerability version identification. \szz{} demonstrates superior capability in accurately identifying affected versions, achieving F1-scores of 94.42\% and 94.74\% across the two well-known OSS datasets, respectively. 
Furthermore, we apply \szz{} to audit the National Vulnerability Database (NVD), which reveals inaccuracies in existing vulnerable version records.
We reported 16 cases to the NVD, all of which have been officially verified and confirmed.
Moreover, we evaluate the effectiveness of six state-of-the-art data-driven vulnerability detection models on early-stage vulnerabilities. The results indicate that detectors trained only on pre-VFC vulnerability samples suffer a significant performance drop when identifying early-stage vulnerabilities. Augmenting the training data with early-stage samples substantially improves detection performance for both initial and pre-VFC vulnerabilities.

\vspace{0.05in}
In summary, the contributions of this paper are as follows:
\begin{itemize}[leftmargin=*]
\item 
%{\bf Methodology:} 
We propose \szz{}, a novel framework that leverages semantic dependency analysis to accurately pinpoint vulnerability-introducing commits from vulnerability-fixing commits, facilitating the collection of early-stage samples for better vulnerability analysis.
\item 
%{\bf Benchmark:} 
We construct a large-scale, manually annotated VIC benchmark, comprising 783 real-world samples curated from recent NVD records and two popular open-source projects (i.e., tcpdump and Vim). Both the datasets and source code for D-SZZ are open-sourced.\footnote{https://github.com/XSecLab-UTD/D-SZZ}
% from the latest NVD records and popular open-source projects, and will release them to enable future research.
\item %{\bf Practical impact:} 
We evaluated the accuracy of vulnerability versions recorded in the NVD and reported 16 erroneous cases, all of which have been officially verified and confirmed.
\item %{\bf Validation \& insight:} 
We conduct a comprehensive evaluation of six representative detectors on early-stage vulnerabilities and assess the impact of incorporating early-stage vulnerabilities into model training on detection performance.
\end{itemize}

%% file: sections/2preliminaries.tex
\section{Preliminaries and Motivation}
% \section{Background}
\label{sec:prelim}

\subsection{Vulnerability-Introducing Commits}
\label{sec:szzbg}
In the lifecycle of a software vulnerability in version control systems such as Git~\cite{git}, the VIC and the VFC represent two key milestones. VIC is a set of code changes that first introduces the vulnerability into the codebase. VFC, also called security patch, is the commit that fixes the vulnerability. Between these two commits, the code remains in a vulnerable state. The VFC is often disclosed or linked in vulnerability advisories (e.g., CVE/NVD entries, GitHub Security Advisories) whereas the VIC is typically not explicitly provided.% and must be inferred through SZZ-style analysis. %Recently, the VIC has shown significance in vulnerability prediction tasks~\cite{le2024latent}. 

The SZZ algorithm is initially proposed to identify bug-introducing commits, and it has subsequently been used to identify VICs, given VFCs as inputs~\cite{fan2019impact}. Specifically, it assumes that the deleted lines in a fixing commit are the cause of the bug~\cite{sliwerski2005changes} and therefore applies change attribution on all of the deleted lines in the fixing commit to detect the bug-inducing commit. 
Later studies attempt to improve the accuracy of SZZ-based algorithms by filtering noise from code changes. For example, AG-SZZ~\cite{kim2006automatic} filters out the addition of blank lines, while MA-SZZ~\cite{da2016framework} avoids blaming meta operations such as renaming. Recent work like NEURALSZZ~\cite{tang2023neural} trains a graph neural network to identify buggy statements for blaming. However, all of the above algorithms can only handle VFCs with deleted lines. SEM-SZZ~\cite{tang2024enhancing} determines buggy statements by comparing control and data flow changes before and after the fixing commit. LLM4SZZ~\cite{tang2025llm4szz} leverages LLMs to analyze code and detect buggy statements. Meanwhile, RA-SZZ~\cite{neto2018impact} and V-SZZ~\cite{bao2022vszz} note that single-step attribution may identify a refactoring commit rather than the actual bug- or vulnerability-introducing commit, so they skip refactoring commits when tracing change history. 
%Over time, the evolution of the SZZ algorithm has diverged significantly from its original version, but we argue that the core idea remains consistent: identifying the statement that introduces a bug or vulnerability as the blaming statement and tracing the commit that introduced its semantics. 
Additionally, although most SZZ algorithms~\cite{neto2018impact, tang2023neural, kim2006automatic, sliwerski2005changes, tang2025llm4szz, tang2024enhancing} are proposed for identifying bug-introducing commits, they are also applied to the vulnerable version detection~\cite{nguyen2016automatic,bao2022vszz}, the just-in-time vulnerability detection~\cite{nguyen2024code}, and the bug bisection~\cite{zhang2025llmbisect}.

\subsection{Vulnerability Detection and Datasets}

Early vulnerability detection tools rely on human-defined rules to perform dynamic or static analysis on target programs~\cite{Cppcheck, Flawfinder, Splint}. However, 
a significant human effort is required to formulate these rules, which limits their scalability. To address this challenge, a series of learning-based methods have been proposed to improve the efficiency. % of vulnerability detection.
Li et al.~\cite{li2021sysevr} apply bidirectional long short-term memory (BLSTM) to predict whether code snippets contain security vulnerabilities, while Russell et al.~\cite{russell2018automated} propose a model based on convolutional neural network (CNN) and random forest. VulDeePecker~\cite{li2018vuldeepecker} uses graph representations of code to improve model performance, then Devign~\cite{zhou2019devign} and ReVeal~\cite{chakraborty2021deep} train graph neural networks (GNNs) to perform vulnerability detection based on code graph representations. With the rise of Transformer models, studies such as LineVul~\cite{fu2022linevul} have begun applying such architectures to vulnerability detection, while more recent works further leverage pre-trained language models~\cite{feng2020codebert, liu2024pre, guo2020graphcodebert, guo2022unixcoder}. %Constrained by the input length of the models, 

Existing approaches primarily focus on detecting vulnerable code snippets at the function or method level, which is also the mainstream granularity level of current vulnerability datasets.
% Current vulnerability datasets can be mainly divided into synthetic vulnerability datasets and natural vulnerability datasets. Among synthetic datasets, SARD~\cite{SARD} and SATE IV~\cite{sate} have generated over 70,000 synthetic vulnerability samples through scripts and manual creation. VulGen~\cite{nong2023vulgen} and VGX~\cite{nong2024vgx} use learning-based methods to insert vulnerabilities into code snippets. However, synthetic data, due to certain differences from human-written code, generally performs worse than natural data in training models.
% For natural datasets, Devign~\cite{zhou2019devign} and Reveal~\cite{chakraborty2021deep} collect vulnerability data based on vulnerability reports from open-source projects. BigVul~\cite{fan2020ac} obtains vulnerability data by reversing patch records from NVD. PrimeVul~\cite{ding2024vulnerability}, building upon multiple natural datasets, filters out multi-file patches to reduce noise in the dataset. However, even for PrimeVul, which combines multiple datasets, or BigVul, which uses all C language patches from NVD, the number of vulnerability samples is only around 10,000, which is insufficient for training models.
%
To train vulnerability detection models, some work~\cite{hanif2022vulberta, wu2022vulcnn, li2021sysevr} adopts synthetic datasets such as SARD~\cite{SARD} and SATE VI~\cite{sate}. Samples in these datasets are directly generated based on known vulnerability patterns, resulting in oversimplified vulnerability features and failing to represent diverse vulnerability contexts. Although VulGen~\cite{nong2023vulgen} and VGX~\cite{nong2024vgx}, attempt to address this by injecting vulnerability-inducing patterns into code snippets from real-world projects. However, such transformations do not always produce valid vulnerabilities, which may introduce noisy samples into the resulting datasets and affect the effectiveness of model training.
Due to the limitations of synthetic datasets, sizable real-world datasets such as Devign~\cite{zhou2019devign}, ReVeal~\cite{chakraborty2021deep}, BigVul~\cite{fan2020ac}, CVEfixes~\cite{bhandari2021cvefixes}, PatchDB~\cite{wang2021patchdb}, CrossVul~\cite{nikitopoulos2021crossvul}, DiverseVul~\cite{chen2023diversevul}, and PrimeVul~\cite{ding2024vulnerability} have been constructed directly from OSS projects, typically by collecting and reversing VFCs. Specifically, VFCs are first collected through open-source databases such as NVD and Bugzilla. Then, changed functions in pre-commit versions are labeled as vulnerable, while post-commit versions and unchanged functions are labeled as non-vulnerable. However, functions may undergo multiple modifications between vulnerability introduction and fixing, so collecting only the fixed-time versions overlooks the semantic changes that arise during vulnerability evolution. As a result, datasets focusing on early-stage vulnerabilities remain limited~\cite{le2024latent}.

% \subsection{Code Property Graph}
% Code Property Graph (CPG) is an abstracted code representation that combines Abstract Syntax Tree (AST), Control Flow Graph (CFG), and Program Dependence Graph (PDG) into a unified data structure~\cite{yamaguchi2014modeling}. AST is a tree-based code representation generated through syntax analysis. CFG is a representation of all paths that might be traversed through a program during its execution. PDG contains the control dependencies and data dependencies of the program. Through CPG, we can easily obtain information about program control flow, data dependencies, and program abstract syntax.

% Recently, CPG has been widely applied in vulnerability detection tasks~\cite{chakraborty2021deep, hin2022linevd, zou2019mu, xiao2020mvp}. Tree-climber~\cite{treeclimber} is a CPG generation package developed with Python and based on Tree-sitter~\cite{treesitter}, capable of extracting code representations that include AST, CFG, and Define-Use Chain (DUC). Compared to other CPG generation tools like Joern~\cite{joern}, Tree-climber is more lightweight and efficient. We modify tree-climber to make it compatible with C++, and build our code analysis module with it.

%\subsection{SZZ Algorithm}

%% file: sections/3motivation.tex
\subsection{Motivating Example}
\label{motivation}
Although the SZZ algorithm has been used to identify VICs~\cite{le2024latent}, it still faces two limitations: (i) it is generally designed to identify bug-introducing commits rather than precise vulnerable statements, and (ii) existing SZZ variants often rely on LLMs or simple code differencing without a principled strategy for pinpointing blame statements, which limits their accuracy.
% the existing SZZ algorithm either relies on LLMs or simple code differencing to locate the blaming statement, resulting in insufficient accuracy.

\begin{figure}[ht]
    \centering
    \includegraphics[width=0.95\linewidth]{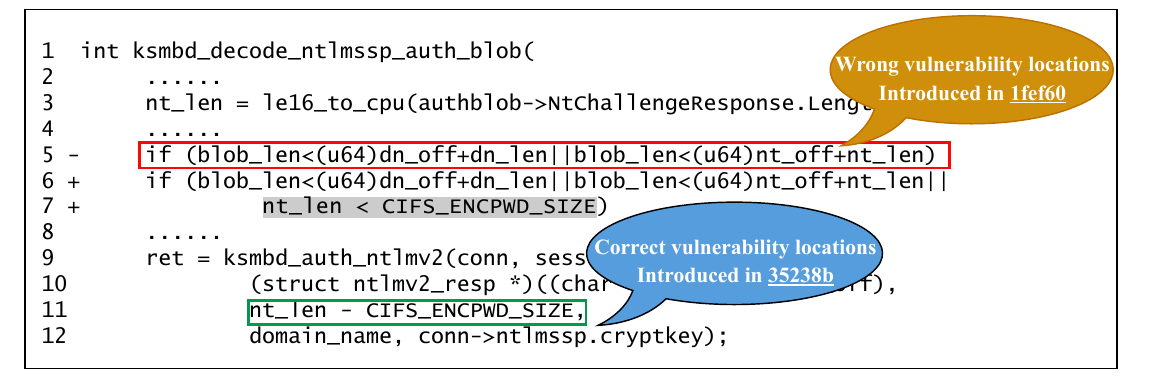}
    \captionsetup{justification=centering}
    \caption{Limitation of SZZ: deleted lines are not always the vulnerable statements}
    \label{fig:motivation_2}
\end{figure}

\autoref{fig:motivation_2} shows an example where the SZZ algorithm fails to locate the precise vulnerable statements in a commit that fixes a heap-overflow vulnerability in \texttt{\small ksmbd}, the kernel SMB service project~\cite{cve20230210}. The fix adds a new condition, \texttt{\small nt\_len < CIFS\_ENCPWD\_SIZE}, to the existing \texttt{\small if} statement that handles abnormal data, thereby preventing negative values from being used in \texttt{\small ret}. In this case, the SZZ algorithm naively treats the deleted \texttt{\small if} statement as the cause of the vulnerability, resulting in the misidentification of the real VIC.

Instead, by analyzing the changes in security patches, we can find that fixes are achieved by checking and updating data states in the vulnerability context. As shown in~\autoref{fig:motivation_2}, the patch filters out abnormal parameters by checking the validity of \texttt{\small nt\_len}. Therefore, the statement that first uses \texttt{\small nt\_len} in the control flow graph downstream of the \texttt{\small if} statement is likely to be the vulnerable statement. This motivates us to propose a data state-aware SZZ algorithm, which can more accurately locate vulnerable statements based on security-fixing patterns and vulnerability scope slicing, thereby improving the overall performance of identifying VICs and enabling the construction of a sizable early vulnerability dataset for studying early-stage vulnerability detection.

%% file: sections/3x5Methodology.tex
\section{Methodology}
\label{sec:method}

\begin{figure*}[ht]
    \centering
    \includegraphics[width=1\linewidth]{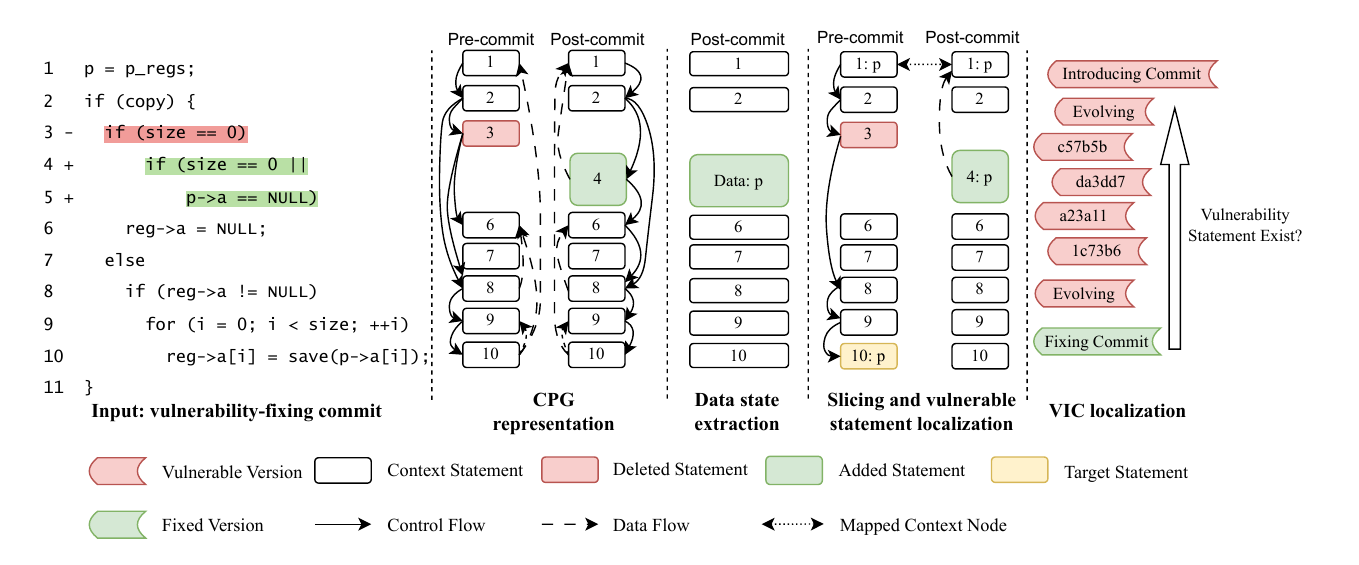}
    \caption{The workflow of \szz{}} 
    \label{fig:overview_szz}
\end{figure*}

%{\color{blue}TBD}
We propose {\underline{d}}ata state-aware {\underline{SZZ}} (\szz{}), a new framework designed to identify VICs by analyzing the statements associated with vulnerable data state.
\autoref{fig:overview_szz} illustrates the workflow of \szz{}. Given a VFC, the system first extracts the CPG for both the pre-VFC and post-VFC code, and then leverages security-fixing patterns to extract vulnerable data states from the post-commit CPG. After that, \szz{} applies vulnerability scope slicing to the code context to pinpoint the specific vulnerable statements interacting with these data states. Finally, it recursively performs change attribution to these vulnerable statements to locate the VIC.

\vspace{0.3em}
\subsection{Vulnerability Scope Slicing}
\subsubsection{Vulnerability-fixing Commit Representation} CPG~\cite{yamaguchi2014modeling}, a code representation that integrates the Abstract Syntax Tree (AST), Control Flow Graph (CFG), and Program Dependency Graph (PDG), enables comprehensive analysis of vulnerable code structures and semantics~\cite{chakraborty2021deep, hin2022linevd, zou2019mu, xiao2020mvp, wang2023graphspd}.  
Each function in a VFC includes two versions: the pre-commit version and the post-commit version, which reflect the vulnerable code before the fix and the patched code after the fix, respectively. 
To establish the connections between these statements and perform vulnerable code analysis, 
we generate two CPGs for each function before and after the commit, respectively. Then, we annotate the CPG nodes using the following three types:
\label{sec:CPG}

\begin{itemize}[leftmargin=*]
    \item \textbf{Deleted nodes}, each corresponding to a deleted statement in the VFC, appear only in the pre-commit CPG and are more likely to contribute to vulnerabilities~\cite{sliwerski2005changes}. 
    \item \textbf{Added nodes}, each corresponding to an added statement in the VFC, appear only in the post-commit CPG. Generally, added lines fix vulnerabilities by adding constraints on the data flow or control flow of the code~\cite{voas1992pie, tang2024enhancing}.
    \item \textbf{Context nodes} represent statements that remain unchanged across the pre- and post-commit versions and therefore appear in both CPGs. Although they are not directly modified by VFCs, they may provide essential context for vulnerability-triggering and fixing statements. %In the CPGs before and after the commit, they can be matched one-to-one.
\end{itemize}
We identify the modified statements and mark the added and deleted nodes in the corresponding post-commit CPG and pre-commit CPG based on their locations (i.e., line number). We then record the relationships of each context node in the both CPGs to establish a cross-version mapping. This allows us to project the vulnerability scope derived from the post-commit CPG analysis back onto the pre-commit version.

\subsubsection{Data State Extraction} 
 
\begin{figure*}[ht]
    \centering

  \begin{subfigure}[b]{0.49\textwidth}
    \centering
    \includegraphics[width=1\linewidth]{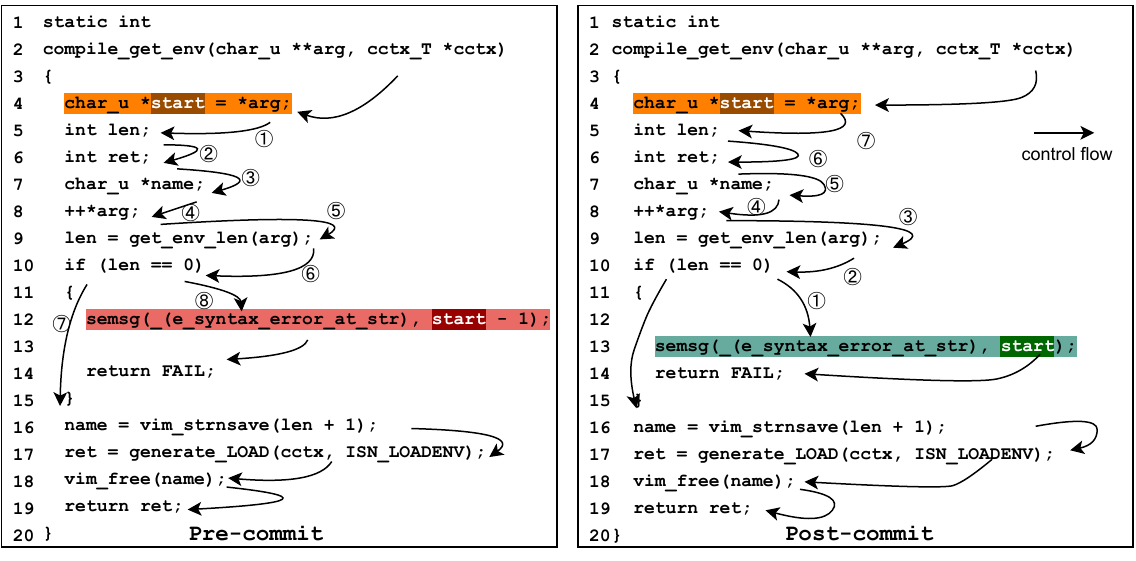}
    \caption{Slicing with both added and deleted lines.}
    \label{fig:Slicing1}
  \end{subfigure}
  \hfill
  \begin{subfigure}[b]{0.49\textwidth}
    \centering
    \includegraphics[width=1\linewidth]{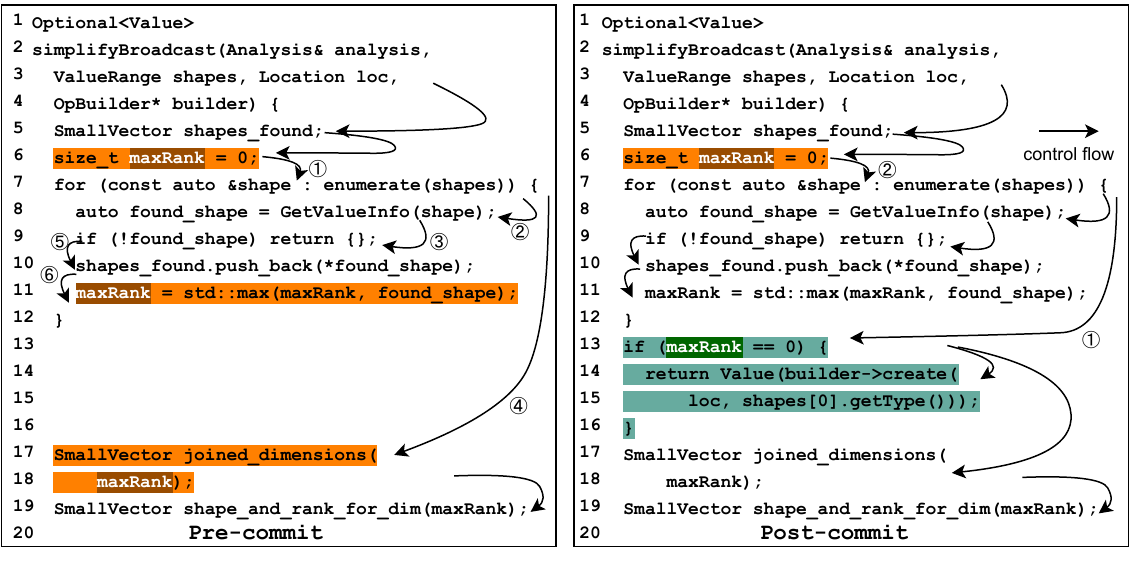}
    \caption{Slicing with only added lines.}
    \label{fig:Slicing2}
  \end{subfigure}

  \caption{Two examples of vulnerability scope slicing}
  \label{fig:Slicing}
\end{figure*}

Security patches typically address vulnerabilities by inserting corrections or validations around the execution point of the abnormal data state~\cite{voas1992pie}. Based on this observation, \szz{} analyzes the vulnerability-fixing actions in a security patch to locate the corrected data states and determine the vulnerable statement. 
Suppose \{$S^{var}$\} denotes the set of all statements of the variable \texttt{\small var}, and $<S^{var}_1$...$S^{var}_n>$ is a statement sequence that conforms to the partial order of the program control flow. During program execution, $S^{var}_i$ executes no later than $S^{var}_{i+1}$. $S^{var}_{add}$ is the added statement in this sequence by the fixing commit, the adjacent statements on the control flow are more likely to be related to the vulnerability, and tracking the semantic introduction points of these statements will help identify the VIC. For example, in~\autoref{fig:Slicing2}, an unexpected value of \texttt{\small maxRank} leads to a security issue in the Application Programming Interface (API) call using \texttt{\small maxRank} at line 17. The patch adds an if statement for security check at line 13, upstream in the control flow of the vulnerable statement, which helps \szz{} identify the vulnerable data state \texttt{\small maxRank}.

Although the recommended best practice is to organize each commit to address a single focused concern~\cite{SurveyBestPractice, 5BestPractices}, such as a bug fix, feature enhancement, or maintenance task, a real-world commit usually entangles a set of code changes to address multiple concerns~\cite{nguyen2013filtering,sun2025dispatch}.
Therefore, in a VFC, not all modified statements and affected data states are related to security fixes. 
To filter out fixing-irrelevant modifications and extract fixing-related data identifiers, we empirically summarize the patterns of security fix behaviors. Specifically, we analyze security patches through iterative sampling to derive fixing patterns, continuing until newly sampled cases reveal no additional patterns. To avoid overlap with the test set, which is constructed from the latter 10\% of CVE records in \autoref{eval}, we perform batch random sampling exclusively from the first 90\%. In this way, we derive the following patterns from 125 sampled cases:
\begin{itemize}[leftmargin=*]
    \item \textbf{Correct value}: Incorrect variable values may affect functionalities such as memory allocation and boundary checks, or directly lead to vulnerabilities like integer overflows. Security fixes typically correct these values by either adding new assignment statements or modifying the right-hand side of existing ones. We extract the left-hand side of the updated assignments as the vulnerable data state.

    \item \textbf{Initialize variable}: Using variables that are declared but not initialized may lead to uninitialized-use vulnerabilities. Security fixes address this by updating the declaration statement to include an initial value for the variable. If a variable declaration lacks initialization in the pre-commit code and the corresponding declaration in the post-commit code includes an initial declarator, we identify the initialized variable as the vulnerable data state. 

    \item \textbf{Update variable type}: Undersized variable type may cause overflows, while oversized type may lead to excessive memory allocation, resulting in denial-of-service. Security fixes modify the type declarations of vulnerable variables. When a change in variable type is introduced by a VFC, we extract the affected variable as the vulnerable data state.

    \item \textbf{Enlarge buffer size}: Enlarging buffer size is a common way to fix buffer overflow. When a VFC modifies the buffer size in a variable declaration or memory allocation, we extract the affected buffer as the vulnerable data state.

    \item \textbf{Add/update concurrency control}: To mitigate race conditions, security fixes add or modify synchronization primitives to regulate access to shared resources. We extract the involved lock variables as the vulnerable data state.

    \item \textbf{Add/update security check}: Security patches fix vulnerabilities such as buffer overflows, null pointer dereferences, and uninitialized references by adding or modifying security checks. We handle the following cases based on the type of modification to the \texttt{\small if} statement: When a complete \texttt{\small if} statement is added, (a)~if the body modifies variable values, we extract the modified variables; (b)~if the body contains jump statements such as \texttt{\small goto} or \texttt{\small return}, we extract all variables referenced in the \texttt{\small if} condition as the vulnerable data state. (c)~When only the conditional expression is modified, we extract the variables involved in the modified condition as the vulnerable data state.

    \item \textbf{Update API call}: Security fixes add or update API calls to address vulnerabilities, either by encapsulating security checks within new API calls or by replacing unsafe operations with secure alternatives. Developers often manipulate the arguments passed to APIs to enforce secure behavior. If a VFC adds a new API call, we extract its argument list as the vulnerable data state. If only the argument list of an existing call is modified, we extract the affected arguments.

\end{itemize}

\noindent Based on the above patterns, we extract vulnerable data state and their statements on the post-commit CPGs.

\subsubsection{Vulnerability Scope Slicing} 
Based on the assumption that security patches tend to validate or rectify nearby data states responsible for vulnerabilities~\cite{voas1992pie}, we employ the statements identified by security-fixing patterns as slicing anchors to enable localized slicing within the procedure. Since these anchors are located in the post-CPG (i.e., the secure version), while the vulnerable statements reside in the pre-CPG (i.e., the vulnerable version), we project the vulnerability scope from the post-CPG to the pre-CPG. To this end, the slicing process consists of the following three steps:

\noindent\textbf{Backward Slicing.} 
We first identify the starting point of the vulnerability scope through backward slicing. Specifically, we use the extracted vulnerable data states and their corresponding statements as anchors, and perform backward slicing to locate where the vulnerable data states originate.
However, performing backward slicing on the CPG has several challenges. First, current static CPG generation tools cannot accurately determine the nature of parameters in API calls, causing data dependencies to tend towards the initial variable definitions. Second, a single node in PDG may depend on a large number of other nodes. Therefore, directly using the CPG for slicing may result in recording irrelevant nodes, leading to unnecessary inspection overhead. 
To mitigate these issues, we adopt an approach inspired by data flow analysis, which tracks the propagation of an assigned variable. Instead of the PDG, we perform backward slicing on the CFG based on identifiers of the data states to locate the relevant nodes.
Specifically, for each extracted data state, we perform a step-by-step backward trace along the control flow until we find the first context node that includes the data identifier. 
For example, in~\autoref{fig:Slicing1}, the security-fixing patterns match a fix involving parameter updates of an API call, which is located in line 13 of the post-commit, with the data state identifier \texttt{\small start} marked in dark green. Subsequently, we perform backward slicing along the control flow and check if the visiting node contains the \texttt{\small start}. After 7 steps, we locate the first context node that defines or uses the variable, which is \texttt{\small char\_u *start = *arg}. This statement, along with the identifier, is recorded as the start of the vulnerability scope.

\noindent\textbf{Statement Mapping.} Next, we project the starting point of the vulnerability scope from the post-CPG to the pre-CPG using the context node mapping established in~\autoref{sec:CPG}.
For instance, in \autoref{fig:Slicing1}, backward slicing yields the context node \texttt{\small char\_u *start = *arg} on line 4, along with the data state \texttt{\small start}. Through the context node mapping, we locate it on line 4 in the pre-commit CPG (marked in orange), which then serves as the anchor for the forward slicing.

\noindent\textbf{Forward Slicing.} We perform forward slicing on the pre-CPG using the context node obtained in the previous step as the anchor. Specifically, we traverse the CFG from the anchor via breadth-first search, checking whether each visited node contains the corresponding data state identifier. If a match is found, we mark it as the vulnerable statement and terminate the slicing. After forward slicing, we obtain the scope of data states on the pre-CPG that are directly affected by the fixing operations. This aligns with our motivation: the fixing commit repairs the vulnerability by validating or correcting the abnormal data states.

In \autoref{fig:Slicing1}, the forward slicing uses the result of statement mapping, i.e., line 4 in the pre-commit CPG, as the anchor. We then slice along the control flow, and after eight steps, we find the first statement that uses the data state \texttt{\small start}, which is deleted by the fixing commit. This statement, marked in red, is then identified as the vulnerable statement. \autoref{fig:Slicing2} illustrates the slicing on the patch that has no deleted lines. After six steps, the algorithm identifies the statements that access the vulnerable data state \texttt{\small maxRank}, which is guarded by the added security check.

\subsubsection{Vulnerability Statement Localization} After performing vulnerability scope slicing, we can identify the statements containing data states that are directly affected by the fixing, which are the statements where forward slicing terminates. We regard these as vulnerable statements and apply change attribution to locate the commits that introduce them. However, we observe that in some cases, the relevant data states do not appear in the forward slicing. There are two possible reasons: (a) There is no downstream statement in the CFG within the current in-function scope that is affected by the patch. For example, the procedure immediately returned after the patch location. (b) The matched variable from the pattern is a global variable, such as a global lock or an environment variable, which prevents us from identifying an affected statement locally. In both cases, the relevant data are beyond the current procedure scope. Therefore, we treat the start of the slice as the vulnerable statement. According to our slicing strategy, this statement can be the function declaration, or the most recent definition or use of the relevant data.

\subsection{VIC Identification} The studies by Neto et al.~\cite{neto2018impact} and Bao et al.~\cite{bao2022vszz} indicate that blaming vulnerable statement with one step cannot accurately identify the VIC. Therefore, when tracing the commit that introduced the vulnerable statement, we check whether the statement exists in the previous commit of the current blame result. If it is,~\szz{} will recursively perform \texttt{\small git blame} on the vulnerable statement until identify the commit that first introduces it or reach the initial commit. Following prior work~\cite{bao2022vszz}, we use edit distance to measure code similarity and perform mapping with a threshold of 0.75.

%% file: sections/4design.tex
\section{VIC Dataset Construction}
\label{sec:dataset}

\begin{table*}[htbp]
\small
\centering
\setlength{\arrayrulewidth}{0.5pt}
\caption{Statistics of VFCs in our constructed VIC datasets}
\begin{tabular}{c|llll}
\noalign{\hrule height 1pt}
Project & \#VFC                   & \# LoC$_{Modified}>$5                  & \#No Del             & Avg LoC$_{Modified}$         \\ \hline 
\multicolumn{1}{c|}{NVD-new} & \multicolumn{1}{c}{557} & \multicolumn{1}{c}{261} & \multicolumn{1}{c}{87} & \multicolumn{1}{c}{56.45} \\ 
\multicolumn{1}{c|}{tcpdump} & \multicolumn{1}{c}{101} & \multicolumn{1}{c}{80} & \multicolumn{1}{c}{1} & \multicolumn{1}{c}{33.88} \\ 
\multicolumn{1}{c|}{Vim} & \multicolumn{1}{c}{125} & \multicolumn{1}{c}{121} & \multicolumn{1}{c}{17} & \multicolumn{1}{c}{20.38} \\ \noalign{\hrule height 1pt}
\end{tabular}

\label{tab:dataset}
\end{table*}

Since prior SZZ studies primarily focus on identifying bug-inducing commits~\cite{lyu2024evaluating, davies2014comparing, wen2019exploring, rosa2021evaluating}, few labeled VIC datasets have been constructed. Although some efforts~\cite{perl2015vccfinder, yang2017vuldigger} collect VIC samples, these datasets are generated through fully automated methods without manual verification. Besides, the existing VIC dataset suffers from several limitations, including small set size, outdated vulnerabilities, limited project variance, and overly simplistic fixes~\cite{bao2022vszz}.

To better evaluate the performance of \szz{} across diverse software projects, we create a manually curated dataset named \textit{NVD-new} from the most recent 10\% of all VFC records (i.e., references labeled as patches) in the NVD snapshot taken in April 2025, which covers 205 OSS repositories from various application domains, including Linux kernel (operating system core), GPAC (multi-media framework), openssl (TLS/SSL library), suricata (intrusion detection system), etc.
To further evaluate the effectiveness of our method on various components of projects with different sizes, we also collect and manually annotate two additional VIC datasets from two popular OSS repositories based on their NVD records: \textit{tcpdump}, a network traffic analyzer with 46K Lines of Code (LoC), and \textit{Vim}, a text editor with 1.2M LoC. Here, we focus on C/C++, the languages with the highest number of vulnerabilities~\cite{c}. \autoref{tab:dataset} summarizes the statistics of VFCs used to construct our VIC datasets.%, including a sufficient number of cases with complex security fixes and samples without deleted lines.

%\subsection{Annotation Method}

% We manually annotate three datasets. The first is a large-scale dataset, where we collect C/C++-related project patches from the National Vulnerability Database (NVD) up to May 2025 and manually analyze the most recent 10\% of them. In addition, we select two popular open-source projects, Vim and tcpdump, and collect all their NVD-recorded patches.

%around 100 samples each. After that, all three annotators discuss the results and derive a codebook based on their discussions. The annotators then label the remaining samples independently. When they encounter data that cannot be clearly classified, they discuss to get agreement and update the codebook accordingly.

Following prior work on identifying how bugs are introduced~\cite{rodriguez2020bugs}, we first analyze the root causes of vulnerabilities, then trace the evolution of the vulnerability context along the commit history, examining each historical commit to assess whether the vulnerability exists, and identify the earliest commit in which the vulnerability appears as the VIC.
%we first analyze the root causes of vulnerabilities using security patches and their descriptions to determine whether a vulnerability exists in the code. We then trace the evolution of the vulnerability context along the commit history, examining each historical commit to assess whether it introduces the vulnerability, and identify the earliest commit in which the vulnerability appears as the VIC.
%
Three authors of this paper, each with over five years of experience in software security, participated in the annotation process. They first analyzed all samples independently and then cross-checked their results. Any disagreements were resolved through discussion until a consensus was reached.
Cases where VFCs are incorrectly labeled by the NVD~\cite{li2017large}, or vulnerabilities are unrelated to C/C++ code are excluded. 
Finally, we obtain three annotated VIC datasets: NVD-new, tcpdump, and Vim, consisting of 557, 101, and 125 samples, respectively. Note that there is no overlap between the NVD new dataset and the Vim and tcpdump datasets.

%% file: sections/5evaluation.tex
\section{Evaluation}

\noindent\textbf{Environment.} We run \szz{} on a server running Ubuntu 22.04, equipped with an Intel W5-2455X CPU (3.50 GHz), 64 GB of RAM, 1 Gbps network bandwidth, and a 1.8 TB NVMe SSD. Model training and evaluation experiments are performed on a computational node featuring two NVIDIA A6000 GPUs (48 GB VRAM each), 32 CPU cores, and 64 GB of RAM.

\noindent\textbf{Implementation.} We generate CPGs using tree-sitter 0.20.1 and tree-climber 0.0.1, extending tree-climber with C++ support. The modified package will be open-sourced along with our project. We use Python scripts to locate vulnerable statements and build the SZZ algorithm on the framework by Bao et al.~\cite{bao2022vszz}. In total, we implement \szz{} with 8.5k LoC.

\label{eval}
% \noindent\textbf{Research questions.} 
\noindent\textbf{Research Questions.} We evaluate \szz{} and explore early-stage vulnerability detection through the following RQs:
\begin{itemize}[leftmargin=*]

\item\textbf{RQ1.} How effectively can \szz{} identify commits that introduce vulnerabilities? 

\item\textbf{RQ2.} How does each main component contribute to the effectiveness of D-SZZ? 

\item\textbf{RQ3.} How useful is \szz{} in helping identify vulnerable versions?

\item\textbf{RQ4.} How well can existing models detect early-stage vulnerabilities in a timely manner?

\item\textbf{RQ5.} What is the overhead of \szz{} to locate early-stage vulnerabilities?

\end{itemize}
% In RQ1 and RQ2, we compare D-SZZ with four SZZ baselines and conduct an ablation study on its main components to demonstrate the novelty of D-SZZ in identifying VICs. In RQ3 and RQ4, we evaluate the benefits of using D-SZZ for two early-stage vulnerability tasks. In RQ5, we analyze the overhead of D-SZZ to illustrate its practicality.

%data collection environment is equipped with a 10 Gbps network card, 1 Gbps outbound bandwidth, and a 1.8 TB NVMe hard drive. 

% All the experiments are conducted
% on a Ubuntu 20.04 server with a Intel Xeon 5122 CPU running
% at 3.60GHz, 512 GB RAM, and 2 NVIDIA GeForce RTX 2080
% Ti GPU cards. The deep learning architecture is built on the
% NVIDIA CUDA Toolkit 11.2 and cuDNN v8.1.

% \subsection{Dataset}

\subsection{Effectiveness of \szz{} in Identifying VICs (RQ1)}

\begin{table*}[ht]
\small
\centering
\caption{Effectiveness comparison of identifying VICs between \szz{} and baseline methods }
\setlength{\arrayrulewidth}{0.5pt}
\label{tab:dszz}
\begin{tabular}{c|ccc|ccc|ccc|ccc}
\noalign{\hrule height 1pt}
\multirow{2}{*}{} & \multicolumn{3}{c|}{NVD-new}                                           & \multicolumn{3}{c|}{tcpdump}                                         & \multicolumn{3}{c|}{Vim}      & \multicolumn{3}{c}{Linux}                                       \\  \cline{2-13} 
                        & \multicolumn{1}{c}{Recall} & \multicolumn{1}{c}{Precision} & F1    & \multicolumn{1}{c}{Recall} & \multicolumn{1}{c}{Precision} & F1    & \multicolumn{1}{c}{Recall} & \multicolumn{1}{c}{Precision} & F1    & \multicolumn{1}{c}{Recall} & \multicolumn{1}{c}{Precision} & F1 \\ \hline
SZZ                    & \multicolumn{1}{c}{42.19}  & \multicolumn{1}{c}{18.45}     & 25.67 & \multicolumn{1}{c}{47.52}  & \multicolumn{1}{c}{21.27}     & 29.39 & \multicolumn{1}{c}{60.80}  & \multicolumn{1}{c}{39.90}     & 48.18 & \multicolumn{1}{c}{54.76}  & \multicolumn{1}{c}{35.06}     & 42.74  \\ 
V-SZZ                    & \multicolumn{1}{c}{59.07}  & \multicolumn{1}{c}{24.41}     & 34.54 & \multicolumn{1}{c}{62.38}  & \multicolumn{1}{c}{29.09}     & 39.68 & \multicolumn{1}{c}{74.40}  & \multicolumn{1}{c}{60.87}     & 66.96   & \multicolumn{1}{c}{53.31}  & \multicolumn{1}{c}{33.76}     & 41.34 \\ 
SEM-SZZ                  & \multicolumn{1}{c}{29.32}  & \multicolumn{1}{c}{40.55}     & 34.03  & \multicolumn{1}{c}{31.68}  & \multicolumn{1}{c}{32.98}     & 32.32 & \multicolumn{1}{c}{53.60}  & \multicolumn{1}{c}{60.00}     & 56.62 & \multicolumn{1}{c}{41.50}  & \multicolumn{1}{c}{45.42}     & 43.37 \\ 
LLM4SZZ                  & \multicolumn{1}{c}{30.88}  & \multicolumn{1}{c}{38.05}     & 34.09 & \multicolumn{1}{c}{22.77}  & \multicolumn{1}{c}{22.54}     & 22.66 & \multicolumn{1}{c}{53.60}       & \multicolumn{1}{c}{55.04}          &    54.31 & \multicolumn{1}{c}{\textcolor{gray}{52.59}}       & \multicolumn{1}{c}{\textbf{\textcolor{gray}{50.26}}}          &    \textcolor{gray}{51.40}   \\  
% CommitShield                 & \multicolumn{1}{c}{20.11}  & \multicolumn{1}{c}{23.48}     & 21.66 & \multicolumn{1}{c}{11.88}  & \multicolumn{1}{c}{10.34}     & 11.06 & \multicolumn{1}{c}{8.82}       & \multicolumn{1}{c}{8.57}          &    8.70 & \multicolumn{1}{c}{\textcolor{gray}{26.86}}       & \multicolumn{1}{c}{\textcolor{gray}{33.33}}          &    \textcolor{gray}{29.75}   \\  
Vercation                & \multicolumn{1}{c}{33.93}  & \multicolumn{1}{c}{\textbf{45.65}}     & 38.93  & \multicolumn{1}{c}{43.56}  & \multicolumn{1}{c}{\textbf{44.00}}     & 43.78 & \multicolumn{1}{c}{72.00}  & \multicolumn{1}{c}{61.64}     & 66.42 & \multicolumn{1}{c}{\textcolor{gray}{19.60}}  & \multicolumn{1}{c}{\textcolor{gray}{43.87}}     & \textcolor{gray}{27.09} \\
LLMBisect                  & \multicolumn{1}{c}{-}  & \multicolumn{1}{c}{-}    &  36.67 & \multicolumn{1}{c}{-}  & \multicolumn{1}{c}{\textbf{-}}    & 48.94  & \multicolumn{1}{c}{-}     & \multicolumn{1}{c}{-}          &    31.25    & \multicolumn{1}{c}{-}       & \multicolumn{1}{c}{-}          &    \textbf{\textcolor{gray}{66.17}} \\ \hline
D-SZZ                    & \multicolumn{1}{c}{\textbf{72.71}}  & \multicolumn{1}{c}{28.78}     & \textbf{41.24} & \multicolumn{1}{c}{\textbf{71.29}}  & \multicolumn{1}{c}{40.45}     & \textbf{51.61} & \multicolumn{1}{c}{\textbf{84.80}}  & \multicolumn{1}{c}{\textbf{68.15}}     & \textbf{75.57} & \multicolumn{1}{c}{\textbf{59.94}}       & \multicolumn{1}{c}{34.21}          &    43.56 \\ \noalign{\hrule height 1pt}
\end{tabular}
\end{table*}

\noindent\textbf{Baseline.} 
To evaluate the effectiveness of \szz{} in identifying commits that first introduce the vulnerability to the repository, we compare it with six existing VIC identification algorithms and adopt the artifacts provided by their original papers:
\begin{itemize}[leftmargin=*]
    \item\textbf{SZZ}~\cite{sliwerski2005changes} is originally developed for locating bug-inducing changes. Previous work~\cite{le2024latent} uses it to identify VIC by directly blaming the deleted lines in the VFC.
    \item\textbf{V-SZZ}~\cite{bao2022vszz} adapts SZZ to refine software versions affected by CVE vulnerabilities and identifies the commit that first introduced the deleted lines in the VFC as the VIC.
    \item\textbf{Sem-SZZ}~\cite{tang2024enhancing} blames statements in fixing commits that affect the code control flow and data flow to determine bug-inducing commits.% by tracing the commits that introduced these statements.
    \item\textbf{LLM4SZZ}~\cite{tang2025llm4szz} adopts LLMs to identify VICs through feeding the contextual information of bug-fixing patches and commit messages to the LLM.%, and the LLM identifies the commits that introduce bugs.
    \item\textbf{Vercation}~\cite{cheng2025vercation} leverages LLM to identify vulnerable statements on the sliced code snippets and employs AST-based code clone detection to identify the VIC.
    \item\textbf{LLMBisect}~\cite{zhang2025llmbisect} uses LLMs to identify VICs based on the commit messages and code modifications.
  
\end{itemize}
    
\noindent We adopt the artifacts provided by their original papers. For the original SZZ and V-SZZ, we use the replication package provided by Bao et al~\cite{bao2022vszz}. For the rest of the methods, we use the replication package provided by their original papers. For LLM-based baselines, we use the LLM backbones specified in their replication packages. Since GPT-4 and o1, used by Vercation and LLMBisect respectively, are no longer available, we use GPT-5-mini as the closest available alternative with stronger reasoning capabilities~\cite{singh2025openai}.

\noindent\textbf{Datasets.}  In addition to the manually annotated dataset described in \autoref{sec:dataset}, we follow LLMBisect~\cite{zhang2025llmbisect} to construct a supplementary dataset from Linux projects, where commit messages with the \texttt{FIX} tag explicitly indicate the corresponding VICs.

\vspace{0.4em}
\noindent\textbf{Metrics.} %{\color{blue}TBD} 
Following previous works~\cite{bao2022vszz, da2016framework, tang2024enhancing, tang2025llm4szz}, we evaluate \szz{} using recall, precision, and F1-score, formulated as follows:
{\small
$$
Recall = \frac{\#Correct VIC }{\#GroundtruthVIC }
$$
\vspace{0.4em}
$$
Precision = \frac{ \#Correct VIC}{\#IdentifiedVIC } 
$$
\vspace{0.4em}
$$
F1 = \frac{2 \times Recall \times Precision}{Recall + Precision}
$$
}

\vspace{0.4em}

\noindent where \textit{\#IdentifiedVIC} refers to the number of all identified VICs, % that are identified by SZZ algorithms. 
\textit{\#CorrectVIC} denotes the number of correctly identified VICs, and % found by the SZZ algorithm. 
\textit{\#GroundtruthVIC} stands for the number of verified, factual VICs. In our manually annotated dataset, each VFC has one corresponding ground truth VIC. Among these metrics, recall represents the likelihood of correctly identifying the VIC corresponding to each VFC. Precision measures how much noise, i.e., incorrect VICs, is introduced when identifying the correct VICs. The F1-score is the harmonic mean of recall and precision, providing a balanced measure of performance.

\noindent\textbf{Results}. \autoref{tab:dszz} reports the performance of \szz{} and six baseline methods. Notably, LLMBisect assigns exactly one VIC to each VFC, making its recall, precision, and F1-score identical. Therefore, we report F1-score as the representative metric of its performance. Across the three manually annotated datasets and Linux dataset, \szz{} achieves higher recall in identifying correct VICs than the best baseline, V-SZZ, by 13.64\%, 8.91\%, 10.40\% and 6.63\%, respectively. For F1-score, \szz{} outperforms by 6.70\%, 11.93\%, and 8.61\% on the newly curated NVD-based dataset and two popular OSS projects. This benefits from the fact that \szz{} does not limit itself to discovering blaming statements from deleted lines, but instead searches for vulnerable data states guided by generalized security fix patterns. 
For example, in \autoref{code:goodcase_3}, the VFC validates \texttt{\small byts} and \texttt{\small idxs} to ensure they are not \texttt{\small null}, preventing operations on an empty case-folded tree. Although directly adding validation to capture abnormal data is a very common fix, SZZ and V-SZZ, which rely on deleted lines to locate vulnerabilities, fail to produce effective outputs. \szz{} overcomes this by identifying the statement that uses \texttt{\small byts} and \texttt{\small idxs} and successfully captures the correct VIC \texttt{\small 9ccfebdd}. 

\vspace{0.4em}
\begin{lstlisting}[caption={VFC 6669de1b in Vim}, label={code:goodcase_3}]
@@ -5585,10 +5585,12 @@ sug_filltree(spellinfo_T *spin, 
     //Bail out if the tree is empty.
     byts = slang->sl_fbyts;
     idxs = slang->sl_fidxs;
@g@+    if (byts == NULL || idxs == NULL)
+	     return FAIL;@g@
     arridx[0] = 0;
     curi[0] = 1;
     wordcount[0] = 0;
     
\end{lstlisting}
% \end{minipage}\hfill
% \begin{minipage}{0.48\linewidth}

\vspace{0.6em}

\begin{lstlisting}[caption={VFC 85b6747a in Vim}, label={code:goodcase_1}]
 @@ -1513,7 +1513,7 @@ init_ccline(int firstc
     ccline.cmdindent=(firstc>0?indent:0);
     // alloc initial ccline.cmdbuff
@r@-    alloc_cmdbuff(exmode_active?250:indent+1);@r@
@g@+    alloc_cmdbuff(indent+50);@g@
     if (ccline.cmdbuff==NULL)
 	    return FAIL;        
\end{lstlisting}
% \end{minipage}

We also observe that SEM-SZZ achieves relatively high precision across all four datasets, but its recall lags significantly. Our analysis attributes this primarily to two limitations. First, SEM-SZZ is designed for bug-inducing commits rather than vulnerability analysis, which prevents it from accurately locating key statements in security fixes. Second, when multiple commit candidates are identified, SEM-SZZ selects only the one nearest to the VFC in the commit history. However, this assumption does not align with vulnerability practice~\cite{bao2022vszz}.

\begin{table*}[ht]
\small
\centering
\caption{Ablation study of \szz{} }
\label{tab:ablation}
\setlength{\arrayrulewidth}{0.5pt}
\setlength{\tabcolsep}{3pt}
% \resizebox{\textwidth}{!}{
\begin{tabular}{c|ccc|ccc|ccc|ccc}
\noalign{\hrule height 1pt}
\multirow{2}{*}{Method} & \multicolumn{3}{c|}{NVD-new}                                           & \multicolumn{3}{c|}{tcpdump}                                         & \multicolumn{3}{c|}{Vim}          & \multicolumn{3}{c}{Linux}                                    \\ \cline{2-13} 
                        & \multicolumn{1}{c}{Recall} & \multicolumn{1}{c}{Precision} & F1    & \multicolumn{1}{c}{Recall} & \multicolumn{1}{c}{Precision} & F1    & \multicolumn{1}{c}{Recall} & \multicolumn{1}{c}{Precision} & F1   & \multicolumn{1}{c}{Recall} & \multicolumn{1}{c}{Precision} & F1 \\ \hline
D-SZZ w/o patterns        & \multicolumn{1}{c}{\textbf{74.51}}       & \multicolumn{1}{c}{9.24}          &   16.44    & \multicolumn{1}{c}{\textbf{72.28}}  & \multicolumn{1}{c}{21.44}     & 33.07 & \multicolumn{1}{c}{84.80}  & \multicolumn{1}{c}{19.03}     & 31.09 & \multicolumn{1}{c}{56.20}  & \multicolumn{1}{c}{10.29}     & 17.39 \\ 
D-SZZ w/ traditional slicing & \multicolumn{1}{c}{67.32}       & \multicolumn{1}{c}{7.44}          &   13.40 & \multicolumn{1}{c}{63.37}  & \multicolumn{1}{c}{16.92}     & 26.71 & \multicolumn{1}{c}{73.60}  & \multicolumn{1}{c}{12.29}     & 21.06 & \multicolumn{1}{c}{56.94}  & \multicolumn{1}{c}{9.09}     & 15.67 \\ 
  D-SZZ w/o slicing          & \multicolumn{1}{c}{63.19}  & \multicolumn{1}{c}{\textbf{28.90}}     & 39.66 & \multicolumn{1}{c}{59.40}  & \multicolumn{1}{c}{34.46}     & 43.62 & \multicolumn{1}{c}{68.80}  & \multicolumn{1}{c}{64.03}     & 66.33  & \multicolumn{1}{c}{47.55}  & \multicolumn{1}{c}{\textbf{36.81}}     & 41.49 \\ \hline
D-SZZ                    & \multicolumn{1}{c}{72.71}  & \multicolumn{1}{c}{28.78}     & \textbf{41.24} & \multicolumn{1}{c}{71.29}  & \multicolumn{1}{c}{\textbf{40.45}}     & \textbf{51.61} & \multicolumn{1}{c}{\textbf{84.80}}  & \multicolumn{1}{c}{\textbf{68.15}}     & \textbf{75.57} & \multicolumn{1}{c}{\textbf{59.94}}       & \multicolumn{1}{c}{34.21}          &    \textbf{43.56} \\ \noalign{\hrule height 1pt}
\end{tabular}
% }
\end{table*}

Another observation is that LLM-based methods are fundamentally limited by their reliance on informative commit messages and sufficient context length, resulting in inconsistent performance across datasets. On the Linux dataset, LLM-based methods exhibit superior performance (highlighted in gray), however, it is likely due to the atypically detailed commit messages. Following LLMBisect~\cite{zhang2025llmbisect}, the Linux VFC commit messages contain the corresponding VIC ID along with a detailed analysis of how and where the vulnerability was introduced, providing additional clues for LLMs. Ordinary patches typically lack such information.
Indeed, LLM-based baselines, LLM4SZZ, Vercation, and LLMBisect, perform suboptimally on all three other datasets. For instance, LLM4SZZ queries the LLM for confidence in candidate VICs and falls back to a rank-based strategy when the LLM responds negatively. Using their original backend (gpt-4o-mini), we find this fallback occurs in 40\% of cases, degrading performance. Similarly, LLMBisect fails on the VFC without informative commit messages, such as \texttt{\small e396648e} with a commit message only "Fixed \#2581". Vercation suffers when the LLM fails to locate vulnerable statements, falling back to using only deleted lines in the VFC.
LLMs also struggle with limited context length. When they require the LLM to assess long code changes, vulnerable statements may be truncated. For example, \autoref{code:goodcase_1} fixes an overflow by enlarging a buffer, \szz{} correctly identifies the deleted \texttt{\small alloc\_cmdbuff} statement and the VIC \texttt{\small 071d4279}. LLM4SZZ identifies the correct vulnerable statement, but when tracing reaches \texttt{\small 4c6fe2e2}, the statement falls outside the context window, leading to an incorrect result.

\subsection{Effectiveness of Main Components (RQ2)} 

To verify the necessity of each main component in \szz{}, we conduct ablation experiments with the above four datasets and analyze the results in \autoref{tab:ablation}. Specifically, in \textit{D-SZZ w/o patterns}, we remove the component that extracts data states based on security-fixing patterns, and treat all patch changes to identify the vulnerable statements. In \textit{D-SZZ w/ traditional slicing}, we use the traditional slicing to locate the vulnerable statement. In \textit{D-SZZ w/o slicing}, we remove the whole vulnerability scope slicing component and only match the data modified by fixing changes within the deleted lines to identify vulnerable statements.

According to the results, D-SZZ w/o patterns achieves slightly higher recall but suffers a significant drop in precision, reaching as low as 9.24\% on the NVD-new dataset. This is expected, since without fixing patterns, the SZZ considers many modifications unrelated to security fixes, thereby introducing noisy commits. We also observe that, even when all modified data states in commit changes are considered, recall increases by only 1.80\% on \textit{NVD-new} dataset. This indicates that the fixing patterns in D-SZZ provide comprehensive coverage.
D-SZZ w/o slicing achieves comparable precision to D-SZZ on the NVD-new and Linux datasets but significantly lower precision on the other two. This suggests that our slicing approach broadens the analysis context to capture more correct vulnerable statements and VICs without introducing noise from the extended scope.
Overall, D-SZZ achieves the best F1-score across all datasets, demonstrating that combining security-fixing patterns with vulnerability scope slicing effectively balances VIC identification capability and the risk of introducing false positives.

%Through the above case studies, we can see that D-SZZ can identify VICs in a broader range of VFCs and achieves higher precision compared to existing SZZ algorithms.
\subsection{Effectiveness in Identifying Vulnerable Versions (RQ3)}
OSS users rely on vulnerable software versions (e.g., Known Affected Software Configurations (KASC) reported by the NVD) to assess their exposure and decide whether to upgrade or apply patches~\cite{kascsite}. Previous works~\cite{nguyen2016automatic, bao2022vszz, cheng2025vercation} have leveraged SZZ to identify vulnerable versions and have reported several inaccuracies in KASC records. Motivated by these findings, this subsection evaluates: (i) the effectiveness of D-SZZ in identifying vulnerable versions, and (ii) potential inconsistencies in KASC records in the NVD.
\vspace{0.3em}
\subsubsection{Effectiveness Comparison on Identifying Vulnerable Versions} 

\begin{table*}[ht]
\small
\centering
\caption{Effectiveness comparison of identifying vulnerable versions between \szz{} and baseline methods}
\label{tab:version}
\setlength{\arrayrulewidth}{0.5pt}
\begin{tabular}{c|ccc|ccc}
\noalign{\hrule height 1pt}
\multirow{2}{*}{Method} & \multicolumn{3}{c|}{tcpdump}                                         & \multicolumn{3}{c}{Vim}                                             \\ \cline{2-7} 
                        & \multicolumn{1}{c}{Recall} & \multicolumn{1}{c}{Precision} & F1    & \multicolumn{1}{c}{Recall} & \multicolumn{1}{c}{Precision} & F1    \\ \hline
SZZ                   & \multicolumn{1}{c}{70.35}  & \multicolumn{1}{c}{97.62}     & 81.77 & \multicolumn{1}{c}{80.86}  & \multicolumn{1}{c}{98.30}     & 88.73 \\ 
V-SZZ                   & \multicolumn{1}{c}{77.11}  & \multicolumn{1}{c}{92.66}     & 84.17 & \multicolumn{1}{c}{84.80}  & \multicolumn{1}{c}{97.82}     & 90.85 \\ 
SEM-SZZ                 & \multicolumn{1}{c}{75.61}  & \multicolumn{1}{c}{96.05}     & 84.61 & \multicolumn{1}{c}{73.19}  & \multicolumn{1}{c}{97.45}     & 83.60 \\

LLM4SZZ                 & \multicolumn{1}{c}{59.41}  & \multicolumn{1}{c}{\textbf{99.13}}     & 74.30 & \multicolumn{1}{c}{74.87}  & \multicolumn{1}{c}{\textbf{98.32}}     & 85.01 \\ 
Vercation                & \multicolumn{1}{c}{82.45}  & \multicolumn{1}{c}{95.54}     & 88.51 & \multicolumn{1}{c}{85.36}  & \multicolumn{1}{c}{86.83}     & 86.09 \\ \hline
D-SZZ                  & \multicolumn{1}{c}{\textbf{94.73}}  & \multicolumn{1}{c}{94.10}     & \textbf{94.42} & \multicolumn{1}{c}{\textbf{92.28}}  & \multicolumn{1}{c}{97.33}     & \textbf{94.74} \\ \noalign{\hrule height 1pt}
\end{tabular}
\end{table*}

\autoref{tab:version} presents the performance of \szz{} and the baselines on identifying vulnerability-affected versions. Since automated version extraction relies on rigorous project management~\cite{imtiaz2023your}, we conduct this experiment on two well-managed open-source projects with manual annotations: tcpdump and Vim. To establish the ground truth, we use \texttt{\small git tag} to collect all tags containing our manually annotated VICs as the affected versions. For \szz{} and other SZZ algorithms, we collect the tags corresponding to all returned VICs and take their union as the predicted version range.

In these two datasets, \szz{} achieves the best overall performance, with its F1-score exceeding the second-ranked baseline by 5.91\% and 3.89\%, respectively, demonstrating its strong capability in identifying the scope of vulnerabilities. We also observe that both SEM-SZZ and LLM4SZZ achieve high precision but significantly lower recall than \szz{} and Vercation. This is because they tend to predict commits closer to the fixing time as the introducing commits, causing them to miss versions that contain vulnerabilities introduced at earlier stages. Vercation achieves precision comparable to \szz{}, but its recall is notably lower, as it falls back to analyzing only deleted lines when the LLM fails. Overall, with its specialized design for vulnerability identification, \szz{} more effectively detects vulnerable versions.
\vspace{0.3em}
\begin{lstlisting}[caption={VIC 2f1714e7 hit by \szz{}}, label={code:index3}]
@@ -718,28 +651,73 @@ rpl_dio_printopt(netdissect_options *ndo,
@r@-   opt = (struct nd_rpl_option *)(((char *)opt) + optlen);@r@
@r@-   length -= optlen; @r@
@r@-   } @r@
@r@-   return; @r@
...
@@ -748,34 +726,67 @@ rpl_dio_print(netdissect_options *ndo
@g@+  opt = (struct rpl_dio_genoption *)(((char *)opt) + optlen);@g@
@g@+  length -= optlen;@g@
@g@+  } @g@ 
@g@+  return;@g@
\end{lstlisting}

When analyzing the results, we find that SZZs achieve better performance on vulnerable version detection than on VIC identification. For example, \szz{} achieves an F1-score of 75.57\% for VIC identification but reaches 94.74\% for vulnerable version detection on Vim dataset. This gap arises from the difference in granularity between the two tasks: a single version may encompass multiple commits, and correctly identifying any one of them is sufficient to determine the affected version.
For instance, in \autoref{code:index3}, \szz{} correctly identifies the vulnerable statement on line 8, but fails to account for a refactoring that occurred during vulnerability evolution, preventing it from tracing back to the correct VIC committed two days earlier. Nevertheless, since both commits are released under the same version number, \szz{} still identifies the correct vulnerable version.
This observation suggests that, although SZZ algorithms are not yet perfect for VIC identification, their accuracy is already sufficient to benefit downstream tasks such as vulnerable version detection.

\subsubsection{Study of Inconsistent Vulnerable Version Records on NVD}
\label{sec:cpe}

%Previous works~\cite{nguyen2016automatic, bao2022vszz} partially disclose the existence of inaccurate KASC records in the NVD using original SZZ on small manually labeled datasets. %Following these works, 
\vspace{0.3em}
We further validate the NVD KASC records using our larger-scale manually labeled dataset (i.e., NVD-new) and the identification results of \szz{}.
Specifically, we first use the manually annotated VICs that can be correctly identified by \szz{} to determine the earliest version tag of each project that contains the given commit. Then, we map the Common Platform Enumeration (CPE) records in KASC to the corresponding version tags to align them with our manual annotations. We exclude a small number of cases where CVEs are reported very recently and have no KASC records.% in the NVD yet, we remove these cases. We also exclude samples in which project version tags differ substantially from the CPE naming convention.

\begin{figure*}[ht]
    \centering

  \begin{subfigure}[b]{0.325\textwidth}
    \includegraphics[width=\linewidth]{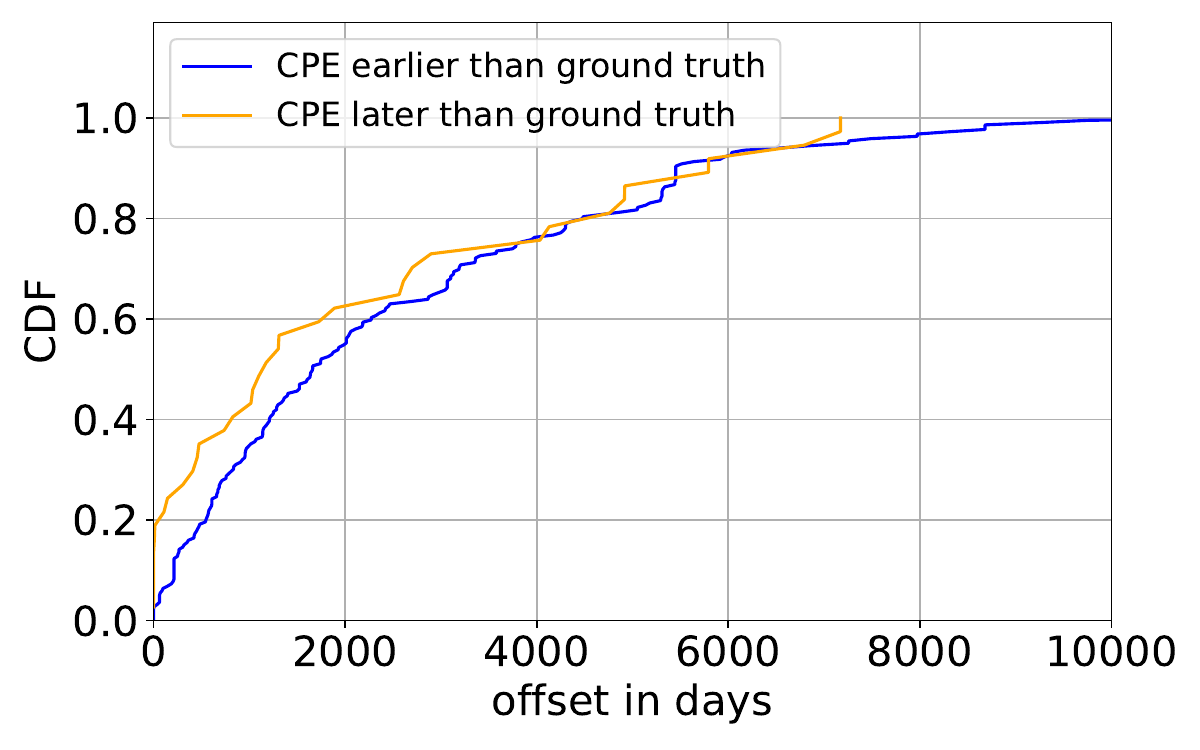}
    \caption{NVD-new}
    \label{fig:CDFtest}
  \end{subfigure}
  \hfill
  \begin{subfigure}[b]{0.325\textwidth}
    \includegraphics[width=\linewidth]{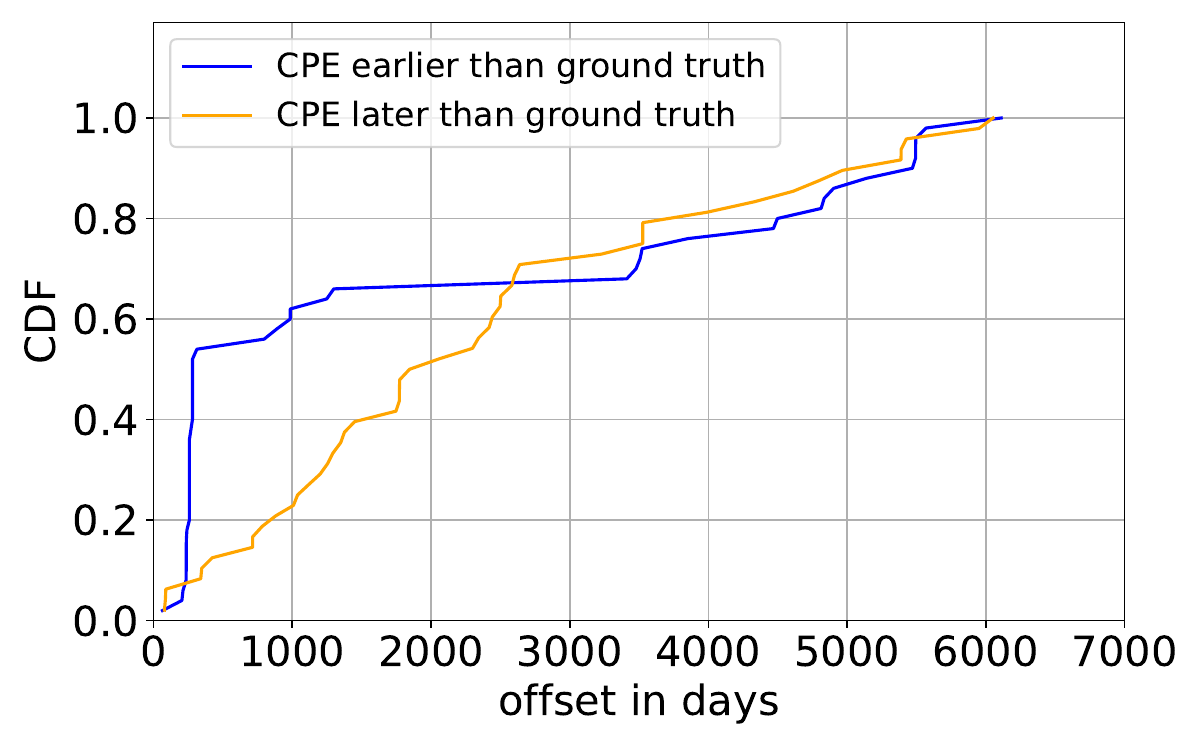}
    \caption{tcpdump}
    \label{fig:CDFtcp}
  \end{subfigure}
  \hfill
  \begin{subfigure}[b]{0.325\textwidth}
    \includegraphics[width=\linewidth]{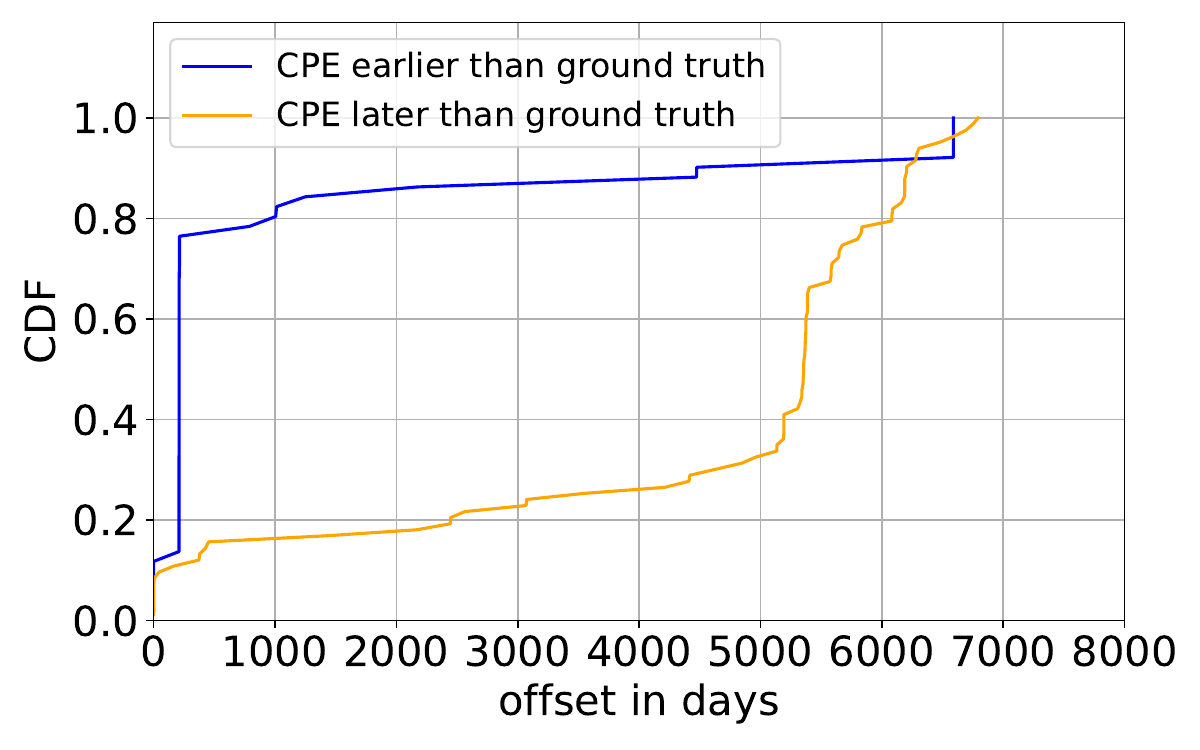}
    \caption{Vim}
    \label{fig:CDFvim}
  \end{subfigure}

  \caption{CDF of the differences in vulnerability start day between NVD records and ground truth}
  \label{fig:CDF}
\end{figure*}

By comparing the start time of vulnerabilities from KASC with those from our manual annotations, we present the results using cumulative distribution functions (CDF) in \autoref{fig:CDF}. In the {NVD-new} dataset, 85\% of KASC records indicate versions earlier than the vulnerable versions identified by manual labeling, with the maximum difference reaching 9,765 days. As shown in the CDF curves, only a small portion of KASC records match or closely align with our manual annotations. This discrepancy may relate to the conservative analysis tendency of NVD analysts: because CVE descriptions and VFC records usually provide only the fixing version, the NVD tends to mark all versions earlier than that as vulnerable~\cite{nguyen2016automatic}. In contrast, datasets from well-known OSS projects benefit from more standardized management, making their KASC records closer to our manual results.

Among these inconsistent results, we report 16 CVEs (from popular OSS projects including the Linux kernel, ImageMagick, and PJSIP) with inaccurate records in the NVD. All of these cases have been officially acknowledged by the NVD, and their affected ranges have been updated. Due to resource constraints, we cannot report every discrepancy to the NVD, as their process requires detailed documentation hosted on a public website with proof-of-concept or in-depth code analysis. Instead, upon acceptance of this paper, we will build a website to release all potential inconsistent versions in our dataset as well as those detected by \szz{}.

\subsection{Assessment of Existing Methods on Early Vulnerability Detection (RQ4)}

Existing learning-based vulnerability detection techniques are typically trained and tested on datasets that use vulnerable code immediately prior to the application of security patches (e.g., BigVul~\cite{fan2020ac}, PatchDB~\cite{wang2021patchdb}, DiverseVul~\cite{chen2023diversevul}, PrimeVul~\cite{ding2024vulnerability}). As a result, their capability to detect early-stage vulnerabilities remains unclear. This subsection investigates: (i) the capability of current learning-based detection techniques to identify early-stage vulnerabilities, and (ii) whether collecting early-stage vulnerabilities using \szz{} can contribute to building more effective detection models.

\begin{table*}[]
\centering
\caption{True positive rate (\%) of existing detectors on original vulnerabilities in VICs}
\small

\setlength{\arrayrulewidth}{0.5pt}
\begin{tabular}{c|cc|cc|cc}
\noalign{\hrule height 1pt}
\centering
\multirow{2}{*}{Test set} & \multicolumn{2}{c|}{GNN} & \multicolumn{2}{c|}{Transformer} & \multicolumn{2}{c}{PLM} \\
                  & Devign     & ReVeal     & VulBERTa        & LineVul       & CodeBERT        & PDBERT        \\ \hline
Pre-VFC vul         & \textbf{8.17}      & \textbf{36.85}      & \textbf{5.90}            & \textbf{63.66}         & 14.42           & 9.78          \\
Original vul       & 4.14       & 29.70      & 5.64            & 3.01          & \textbf{16.54}           & \textbf{12.41}   \\     \noalign{\hrule height 1pt}
\end{tabular}

\label{tab:rq3tp}
\end{table*}

\vspace{0.3em}
\subsubsection{Existing Model Effectiveness in Early Vulnerability Detection} We examine six representative vulnerability detection models across different architectures to evaluate their effectiveness in detecting early-stage vulnerabilities: 
\textit{\textbf{Devign}}~\cite{zhou2019devign} and \textit{\textbf{Reveal}}~\cite{chakraborty2021deep} extract code structural information using CPGs and adopt \textit{GNNs} for vulnerability detection. 
\textit{\textbf{LineVul}}~\cite{fu2022linevul} and \textit{\textbf{VulBERTa}}~\cite{hanif2022vulberta} train BERT/RoBERTa-based \textit{Transformer} models %from scratch based on the BERT and RoBERTa architectures, respectively, and 
for vulnerability detection based on sequence of code tokens.
\textit{\textbf{CodeBERT}}~\cite{feng2020codebert} and \textit{\textbf{PDBERT}}~\cite{liu2024pre} fine-tune \textit{pre-trained language models (PLMs)} with code-specific pretraining tasks to predict software vulnerabilities. We employ the original implementation provided in their papers and the replication package in CodeXGLUE~\cite{lu2021codexglue}.

\vspace{0.3em}
\noindent \textbf{Datasets.} For the training set, we use VFCs from the earliest 90\% of NVD records and adopt the same labeling scheme as prior works~\cite{fan2020ac}, which labels pre-VFC functions (code immediately prior VFCs) as vulnerable and all other functions as non-vulnerable. This yields a training set of 6,497 vulnerable samples and 89,240 non-vulnerable samples.
Previous works evaluate models trained on such datasets by testing on vulnerable functions just before the VFC, referred to as \textit{pre-VFC vulnerability}. To further assess these models on early vulnerability detection, we include \textit{original vulnerability} samples, which are functions where vulnerabilities are first introduced into the codebase. These samples are derived from our annotated VIC datasets based on the latest 10\% of NVD records. If a pre-VFC sample is identical to its original version, we include it only in the pre-VFC set to avoid duplication. As a result, the test set contains 1,186 pre-VFC vulnerability samples and 266 original vulnerability samples.

\vspace{0.3em}
\noindent \textbf{Results.} 
\autoref{tab:rq3tp} presents a comparison of model recall for detecting pre-VFC and original vulnerabilities. We observe that GNN and Transformer-based models perform well on pre-VFC vulnerabilities but suffer a significant drop on the original ones. For example, LineVul achieves a recall of 63.66\% on the pre-VFC but only 3.01\% on original vulnerabilities. This indicates that training solely on pre-VFC data does not enable these models to capture the features of early-stage vulnerabilities.
In contrast, the two pretrained models achieve comparable performance on both pre-VFC and original vulnerabilities. We attribute this to their pretraining on large-scale code corpora, which allows them to better generalize pre-VFC features to the detection of early-stage vulnerabilities.
In summary, pretrained models exhibit consistent performance across different vulnerability stages after fine-tuning, while models trained from scratch, although effective at identifying pre-VFC vulnerabilities, struggle to detect original vulnerabilities.

\vspace{0.3em}
\subsubsection{Impact of Early-Stage Samples on Vulnerability Detection}
The above experimental results indicate that existing models may struggle with original vulnerabilities not seen during training. We further investigate the impact of incorporating early-stage vulnerability samples into the model training.

\vspace{0.3em}
\noindent \textbf{Datasets.} In addition to the above pre-VFC training set based on the earliest 90\% of NVD records, we construct an extended training set of \textit{early-stage vulnerability} samples, including both \textit{original} vulnerabilities, which refer to the vulnerable code at the VIC, and \textit{evolving} vulnerabilities, which refer to the intermediate states between the original and pre-VFC samples. Specifically, for each VFC, we apply \szz{} to identify the VICs and collect the original vulnerability samples, then iteratively traverse the evolution from the VIC to the VFC to collect evolving vulnerability samples. We train the selected models on a training set that includes both pre-VFC and early-stage vulnerability samples from the earliest 90\% of NVD records, and test them on pre-VFC and original vulnerabilities from the rest 10\% of NVD records to evaluate whether incorporating early-stage samples can help improve early vulnerability detection.

\begin{table*}[]
\caption{Impact of incorporating early-stage samples identified by \szz{} into training}
\small
\centering

\setlength{\arrayrulewidth}{0.5pt}

\begin{tabular}{cc|c|ccccc|cc}
\noalign{\hrule height 1pt}
\multicolumn{2}{c|}{Model}                                                    & Training set     & Acc   & B-Acc & F1    & Precision & Recall & Recall$_{pre\_VFC}$ & Recall$_{original}$ \\ \noalign{\hrule height 1pt}
\multicolumn{1}{c|}{\multirow{4}{*}{GNN}}         & \multirow{2}{*}{Devign}   & pre-VFC   & \textbf{88.97} & \textbf{50.67} & 7.16  & \textbf{6.92}      & 7.43   & 8.17               & 4.14             \\  
\multicolumn{1}{c|}{}                             &                           & +early-stage & 79.83 & 49.78 & \textbf{8.25}  & 5.58      & \textbf{15.84}  & \textbf{14.66}              & \textbf{16.10}            \\ \cline{2-10} 
\multicolumn{1}{c|}{}                             & \multirow{2}{*}{ReVeal}   & pre-VFC   & \textbf{95.27} & 67.22 & \textbf{46.30} & \textbf{66.41}     & 35.54  & 36.85              & 29.70            \\  
\multicolumn{1}{c|}{}                             &                           & +early-stage & 91.10 & \textbf{69.99} & 37.25 & 31.24     & \textbf{46.14}  & \textbf{47.55}              & \textbf{39.85}            \\ \hline
\multicolumn{1}{c|}{\multirow{4}{*}{Transformer}} & \multirow{2}{*}{VulBERTa} & pre-VFC   & \textbf{93.47} & \textbf{52.32} & 9.32  & \textbf{22.85}     & 5.85   & 5.90               & 5.64             \\  
\multicolumn{1}{c|}{}                             &                           & +early-stage & 84.61 & 52.24 & \textbf{10.31} & 8.29      & \textbf{13.64}  & \textbf{11.89}              & \textbf{21.43}            \\ \cline{2-10} 
\multicolumn{1}{c|}{}                             & \multirow{2}{*}{LineVul}  & pre-VFC   & \textbf{97.06} & \textbf{76.15} & \textbf{67.17} & \textbf{93.05}     & 52.55  & 63.66              & 3.01             \\  
\multicolumn{1}{c|}{}                             &                           & +early-stage & 90.37 & 75.23 & 40.88 & 31.53     & \textbf{58.1}3  & \textbf{65.77}              & \textbf{24.06}            \\ \hline
\multicolumn{1}{c|}{\multirow{4}{*}{PLM}} & \multirow{2}{*}{CodeBERT} & pre-VFC   & \textbf{91.94} & 55.72 & \textbf{17.39} & \textbf{21.06}     & 14.81  & 14.42              & 16.54            \\  
\multicolumn{1}{c|}{}                             &                           & +early-stage & 82.83 & \textbf{58.13} & 16.78 & 11.62     & \textbf{30.23}  & \textbf{28.33}              & \textbf{38.72}            \\ \cline{2-10} 
\multicolumn{1}{c|}{}                             & \multirow{2}{*}{PDBERT}   & pre-VFC   & \textbf{93.32} & 54.31 & \textbf{14.96} & \textbf{27.59}     & 10.26  & 9.78               & 12.41            \\  
\multicolumn{1}{c|}{}                             &                           & +early-stage & 84.63 & \textbf{54.36} & 13.07 & 9.67      & \textbf{20.18}  & \textbf{17.96}              & \textbf{30.08}            \\ \noalign{\hrule height 1pt}
\end{tabular}
\label{tab:rq3manually}
\end{table*}

\vspace{0.3em}
\noindent \textbf{Results.} \autoref{tab:rq3manually} presents the impact of adding early-stage vulnerability samples to the training set of vulnerability detection models. We find that for all models, incorporating early-stage vulnerability samples not only improves their ability to detect early-stage vulnerabilities but also enhances their performance in identifying pre-VFC vulnerabilities. After adding early-stage samples, all models achieve higher recall, with Devign, VulBERTa, CodeBERT, and PDBERT being enhanced to almost twice their baseline. However, we also observe that adding early-stage samples leads to a decrease in precision. %Balancing false positives and false negatives is an important consideration for VD models in practice. 
Considering both balanced accuracy and F1-scores, we find that most models achieve overall performance comparable to the baseline, suggesting that they may benefit from training with early-stage vulnerability samples.

\subsection{Overhead of \szz{} (RQ5)}
The overhead of \szz{} primarily stems from network bandwidth, disk I/O capacity, and the scale of data. When identifying VICs and extracting evolving vulnerabilities, \szz{} traverses multiple vulnerable software versions within a locally cloned repository. Operations such as \texttt{\small git clone}, \texttt{\small blame}, and \texttt{\small checkout} consume network and disk resources. As a result, the hardware capabilities will affect \szz{}’s performance overhead. %, its overhead is also affected by the total number of VFCs.
%To evaluate \szz{}'s capability to collect evolving vulnerability samples at scale, we use 
Using 5,993 VFCs across 424 software repositories derived from NVD records as input, \szz{} identifies 5,107 VICs within 23.7 hours. On average, for each VFC, 14.24 seconds is taken to retrieve the VIC, which is acceptable for practical use. In contrast, LLMBisect, Vercation, and LLM4SZZ consume 727K, 430K, and 70K tokens, respectively, for the analysis of a single VFC, resulting in lower efficiency than \szz{} in both cost and time.

%% file: sections/8discussion.tex
\section{Discussion}%{\color{blue}merge}
\label{sec:discussion}

Our approach builds on fixing patterns and CPG analysis, currently focuses on identifying VICs in C/C++, which are among the languages with the largest number of reported software vulnerabilities~\cite{vulnc}. As analysis tools such as tree-sitter~\cite{treesitter} continue to broaden language support, we expect our method to be readily extensible to additional languages. Additionally, VIC localization in our approach relies on the version control capabilities provided by Git, the most widely used system for software maintenance~\cite{git}.

Because \szz{} localizes vulnerable statements through static analysis, it can complement LLM-based approaches. Future LLM-based methods can use \szz{} to narrow the search space to the localized vulnerable statements, which reduces analysis costs and improves robustness on commits with uninformative messages. For statement tracing, \szz{} adopts the 0.75 edit-distance threshold from V-SZZ~\cite{bao2022vszz} to stay consistent with prior work, and we leave more adaptive tracing strategies to future work.

The performance of D-SZZ depends on the quality of the input VFCs. First, VFCs may contain incorrect labels~\cite{wpatch}. We use the NVD~\cite{nvdsite} as our data source, where VFCs undergo manual review by both software vendors and the NVD, which effectively mitigates the impact of mislabeled cases. Second, developers may combine security patches with functional changes, introducing noise into VIC identification~\cite{li2024cleanvul}. D-SZZ mitigates this through its patterns for identifying security-fixing behaviors. Moreover, the community has increasingly recognized this challenge~\cite{sun2025dispatch, li2025they}, and as this line of research progresses, D-SZZ stands to benefit from these advances.

%% file: sections/9conclusion.tex
\section{Related Work}
\label{sec:related}

\noindent\textbf{Early-Stage Vulnerability Detection.} 
Most existing learning-based vulnerability detection techniques~\cite{zhou2019devign, fu2022linevul, wu2022vulcnn, hanif2022vulberta} do not consider the detection of early-stage vulnerabilities, as their training and test datasets~\cite{fan2020ac, ding2024vulnerability, chen2023diversevul, bhandari2021cvefixes} are extracted from VFCs and do not contain early-stage vulnerability samples. Le et al.~\cite{le2024latent} extend the BigVul~\cite{fan2020ac} dataset using V-SZZ~\cite{bao2022vszz} and investigate whether incorporating early-stage vulnerabilities can improve the performance of vulnerability detection models. However, the limited effectiveness of SZZ introduces noise into the evaluation. The community has recognized the importance of high-quality early-stage vulnerability datasets for vulnerability detection. V-SZZ~\cite{bao2022vszz} and NeuralSZZ~\cite{tang2023neural} each provide small-scale manually annotated datasets, but they suffer from limitations such as small scale and simplistic samples. LLMBisect~\cite{zhang2025llmbisect} and CommitShield~\cite{wu2025commitshield} leverage LLMs to identify VICs, but they face limitations when commit messages and vulnerability descriptions provide insufficient information. Just-in-time vulnerability detection~\cite{nguyen2024code, nguyen2025toward, kamei2012large} can directly identify commits that introduce vulnerabilities. However, training JIT models themselves requires high-quality, large-scale early-stage vulnerability datasets, creating a circular dependency. Our \szz{} improves the effectiveness of SZZ while providing three high-quality VIC datasets, laying the foundation for future research.

\vspace{0.3em}
\noindent\textbf{Vulnerable Version Identification.} 
Vulnerable version identification helps the OSS users determine whether the open-source software they use is affected by known vulnerabilities. NVD~\cite{nvdsite} employs KASC to record affected CPEs. However, manual analysis can be delayed and has been reported in several studies to contain numerous errors~\cite{nguyen2016automatic, bao2022vszz}. A number of studies attempt to automate the identification of affected versions. Dong et al.~\cite{dong2019towards} use Named Entity Recognition (NER) to analyze vulnerability reports, while Dai et al.~\cite{dai2021facilitating} leverage proof-of-concept exploits and fuzz testing to identify affected versions. Vercation~\cite{cheng2025vercation} identifies the vulnerable version ranges by mapping the VIC and VFC to version tags recorded in software repository. Among existing techniques, combination of the SZZ algorithm~\cite{nguyen2016automatic, bao2022vszz} and code fingerprint-based clone detection~\cite{kim2017vuddy, wu2024vision, woo2021v0finder} shows considerable potential~\cite{cheng2025vercation}.

\vspace{0.3em}
\noindent\textbf{Patch Detangling.} 
Researchers have recognized that entangled commits pose challenges for the patch analysis tasks~\cite{ding2024vulnerability, kim2006automatic, tang2023neural}. To address this issue, a series of approaches have been proposed to decompose tangled commits. Some of them leverage manual or interactive methods to decompose commits~\cite{shen2021smartcommit, safwan2019decomposing, herbold2020large}, while others employ heuristic-based techniques to extract commits for specific purposes~\cite{kirinuki2014hey, barnett2015helping}. Besides, graph-based code representation has been adopted to classify different components within the complex commit, includes program slicing techniques~\cite{guo2017interactively, sun2025dispatch} and learning-based graph node classification approaches~\cite{falleri2014fine, li2022utango, nguyen2019graph}. These advances in commit decomposition are complementary to our \szz{}. By disentangling security-relevant changes from unrelated modifications, \szz{} can benefit from cleaner inputs and achieve more accurate results.

% \subsection{Vulnerability Bisection} 

% \section{Related Work}

\section{Conclusion}
\label{sec:conclusion}
We propose \szz~ that captures structural information of the vulnerable statements through code property graph and security-fixing patterns from vulnerability-fixing commits, and tracks the evolving process of vulnerability semantics to identify vulnerability-introducing commits.
%and determine the vulnerability introduction time. 
Comparison evaluation on large-scale, manually annotated datasets based on the latest NVD records and popular OSS shows that \szz{} substantially outperforms existing SZZ-based approaches. We also reveal the substantial presence of inaccurate vulnerable version records in the NVD, which can significantly delay timely updates for software users. Moreover, our study highlights the limitations of existing detection models on early-stage vulnerabilities, providing valuable insights for future research.

\section*{Acknowledgment}
This research is partially supported by the National Science Foundation (NSF) grants CNS-2450602, CNS-2544345, and DGE-2623088. Any opinions, findings, and conclusions or recommendations expressed in this material are those of the author(s) and do not necessarily reflect the views of the NSF.

% , \szz{} achieves superior performance in identifying vulnerability-introducing commits. In addition, we construct three manually annotated datasets and evaluate the effectiveness of \szz{} in tasks such as vulnerability version identification and early-stage vulnerability detection. Experimental results demonstrate that these tasks benefit from both \szz{} and the early-stage vulnerability data it identifies.
% \szz{} overcomes the limitation of being unable to parse VFCs without deleted lines and improves the precision. Subsequently, \dataset~tracks function changes between these two time points and mines potential vulnerability samples. The evolving vulnerability samples discovered by \dataset~provide richer examples for vulnerability prediction models. Our evaluation reveals that on both manually annotated small datasets and automatically collected large-scale datasets, \dataset~helps models discover more evolving vulnerabilities, contributing to earlier vulnerability detection and fix.

%% file: sections/10dataavailibility.tex
% \section*{Open Science} 

% We are committed to fully following the open science policy of the CCS 2026 research track. The source code, experimental results, documentation of \szz{}, and its related dataset have been made available at \url{https://osf.io/cwjqg/overview?view_only=01a506fc6ffa4dd4914f7e0362df2d88}. We will
% make this artifact publicly available and submit our data and tools to the CCS 2026 artifact evaluation committee upon acceptance of
% our paper.
\section*{Ethical Considerations}
All vulnerabilities studied in this work are collected from public datasets, and the corresponding software vendors have already released security patches or workarounds. Therefore, our study does not introduce new risks related to vulnerability disclosure. In addition, the code snippets, models, and tools used in our study are obtained from open-source projects or public datasets. We follow the relevant license terms during their use to avoid harming to the rights and interests of these entities.

%% file: sections/appendix.tex
\section{Performance on different commit scales}
\label{smallcommit}

\begin{table}[ht]
\centering
\caption{Effectiveness across different commit scales}
\label{tab:small}
\setlength{\arrayrulewidth}{0.5pt}
\resizebox{240 pt}{!}{
\begin{tabular}{c|ccc|ccc}
\noalign{\hrule height 1pt}
\multirow{2}{*}{} & \multicolumn{3}{c|}{small}                                           & \multicolumn{3}{c}{large}                                           \\ \cline{2-7} 
                  & \multicolumn{1}{c}{Recall} & \multicolumn{1}{c}{Precision} & F1    & \multicolumn{1}{c}{Recall} & \multicolumn{1}{c}{Precision} & F1    \\ \hline
SZZ               & \multicolumn{1}{c}{53.11}  & \multicolumn{1}{c}{31.92}     & 39.87 & \multicolumn{1}{c}{40.99}  & \multicolumn{1}{c}{12.26}     & 18.87 \\ 
V-SZZ             & \multicolumn{1}{c}{\textbf{73.03}}  & \multicolumn{1}{c}{43.56}     & 54.57 & \multicolumn{1}{c}{58.62}  & \multicolumn{1}{c}{16.20}     & 25.39 \\ 
SEM-SZZ           & \multicolumn{1}{c}{33.10}  & \multicolumn{1}{c}{46.45}     & 38.66 & \multicolumn{1}{c}{25.00}  & \multicolumn{1}{c}{\textbf{34.03}}     & 28.82 \\ 
LLM4SZZ           & \multicolumn{1}{c}{37.34}  & \multicolumn{1}{c}{53.57}     & 44.01 & \multicolumn{1}{c}{26.81}  & \multicolumn{1}{c}{27.56}     & 27.18 \\ \hline
\szz{}              & \multicolumn{1}{c}{72.19}  & \multicolumn{1}{c}{\textbf{46.52}}     & \textbf{56.59} & \multicolumn{1}{c}{\textbf{73.56}}  & \multicolumn{1}{c}{20.38}     & \textbf{31.92} \\ \noalign{\hrule height 1pt}
\end{tabular}
}
\end{table}

In this section, we compare the performance of ~\szz{} and baselines on commits of two different scales. We split samples from the NVD-new dataset from \autoref{sec:dataset} that contain less than five modified lines as the small set~\cite{bao2022vszz}, and use the remaining samples as the large set. According to the results shown in \autoref{tab:small},~\szz{} achieves the best overall performance on VFCs of both scales. We also observe that although all SZZ algorithms experience performance degradation when handling complex VFCs,~\szz{} maintains a good level of recall. This indicates that the security patterns and slicing used in~\szz{} can comprehensively cover fix types and demonstrate good generalization ability.

% \section{Dataset distribution}

%% file: sample-base.bib
@String{Computing = "Computing" }

@String{Computer = "{IEEE} Computer" }

@String{Springer = "Springer-Verlag" }

@inproceedings{wu2025commitshield,
  title={COmmitSHield: Tracking vulnerability introduction and fix in version control systems},
  author={Wu, Zhaonan and Zhao, Yanjie and Wei, Chen and Wan, Zirui and Liu, Yue and Wang, Haoyu},
  booktitle={2025 IEEE/ACM 47th International Conference on Software Engineering: Companion Proceedings (ICSE-Companion)},
  pages={279--290},
  year={2025},
  organization={IEEE}
}

@article{singh2025openai,
  title={Openai gpt-5 system card},
  author={Singh, Aaditya and Fry, Adam and Perelman, Adam and Tart, Adam and Ganesh, Adi and El-Kishky, Ahmed and McLaughlin, Aidan and Low, Aiden and Ostrow, AJ and Ananthram, Akhila and others},
  journal={arXiv preprint arXiv:2601.03267},
  year={2025}
}

@article{li2024cleanvul,
  title={Cleanvul: Automatic function-level vulnerability detection in code commits using llm heuristics},
  author={Li, Yikun and Zhang, Ting and Widyasari, Ratnadira and Tun, Yan Naing and Nguyen, Huu Hung and Bui, Tan and Irsan, Ivana Clairine and Cheng, Yiran and Lan, Xiang and Ang, Han Wei and others},
  journal={arXiv preprint arXiv:2411.17274},
  year={2024}
}

@article{li2025they,
  title={What Do They Fix? LLM-Aided Categorization of Security Patches for Critical Memory Bugs},
  author={Li, Xingyu and Pu, Juefei and Wu, Yifan and Zou, Xiaochen and Zhu, Shitong and Wu, Qiushi and Zhang, Zheng and Hsu, Joshua and Dong, Yue and Qian, Zhiyun and others},
  journal={arXiv preprint arXiv:2509.22796},
  year={2025}
}

@misc{vulnc,
  title = {What Is the Most Secure Coding Language? Top Options in 2026},
  key = {What Is the Most Secure Coding Language? Top Options in 2026},
  howpublished = {https://www.securityjourney.com/post/what-is-the-most-secure-coding-language},
  year = {2025}
}

@misc{wpatch,
  title = {Heap-based Buffer Overflow in function ins\_compl\_infercase\_gettext() in vim/vim},
  key = {CVE-2022-2522},
  howpublished = {https://huntr.com/bounties/3a2d83af-9542-4d93-8784-98b115135a22
},
  year = {2022}
}

@misc{cve20230210,
  title = {NVD - CVE-2023-0210},
  key = {CVE-2023-0210},
  howpublished = {https://nvd.nist.gov/vuln/detail/CVE-2023-0210},
  year = {2023}
}

@article{zhang2025llmbisect,
  title={LLMBisect: Breaking Barriers in Bug Bisection with A Comparative Analysis Pipeline},
  author={Zhang, Zheng and Li, Haonan and Li, Xingyu and Zhang, Hang and Qian, Zhiyun},
  journal={arXiv preprint arXiv:2510.26086},
  year={2025}
}

@inproceedings{nguyen2019graph,
  title={Graph-based mining of in-the-wild, fine-grained, semantic code change patterns},
  author={Nguyen, Hoan Anh and Nguyen, Tien N and Dig, Danny and Nguyen, Son and Tran, Hieu and Hilton, Michael},
  booktitle={2019 IEEE/ACM 41st International Conference on Software Engineering (ICSE)},
  pages={819--830},
  year={2019},
  organization={IEEE}
}

@inproceedings{li2022utango,
  title={UTANGO: untangling commits with context-aware, graph-based, code change clustering learning model},
  author={Li, Yi and Wang, Shaohua and Nguyen, Tien N},
  booktitle={Proceedings of the 30th ACM Joint European Software Engineering Conference and Symposium on the Foundations of Software Engineering},
  pages={221--232},
  year={2022}
}

@inproceedings{falleri2014fine,
  title={Fine-grained and accurate source code differencing},
  author={Falleri, Jean-R{\'e}my and Morandat, Flor{\'e}al and Blanc, Xavier and Martinez, Matias and Monperrus, Martin},
  booktitle={Proceedings of the 29th ACM/IEEE international conference on Automated software engineering},
  pages={313--324},
  year={2014}
}

@inproceedings{barnett2015helping,
  title={Helping developers help themselves: Automatic decomposition of code review changesets},
  author={Barnett, Mike and Bird, Christian and Brunet, Jo{\~a}o and Lahiri, Shuvendu K},
  booktitle={2015 IEEE/ACM 37th IEEE International Conference on Software Engineering},
  volume={1},
  pages={134--144},
  year={2015},
  organization={IEEE}
}

@inproceedings{kirinuki2014hey,
  title={Hey! are you committing tangled changes?},
  author={Kirinuki, Hiroyuki and Higo, Yoshiki and Hotta, Keisuke and Kusumoto, Shinji},
  booktitle={Proceedings of the 22nd International Conference on Program Comprehension},
  pages={262--265},
  year={2014}
}

@inproceedings{guo2017interactively,
  title={Interactively decomposing composite changes to support code review and regression testing},
  author={Guo, Bo and Song, Myoungkyu},
  booktitle={2017 IEEE 41st annual computer software and applications conference (COMPSAC)},
  volume={1},
  pages={118--127},
  year={2017},
  organization={IEEE}
}

@inproceedings{herbold2020large,
  title={Large-scale manual validation of bugfixing changes},
  author={Herbold, Steffen and Trautsch, Alexander and Ledel, Benjamin},
  booktitle={Proceedings of the 17th International Conference on Mining Software Repositories},
  pages={611--614},
  year={2020}
}

@inproceedings{safwan2019decomposing,
  title={Decomposing the rationale of code commits: the software developer’s perspective},
  author={Safwan, Khadijah Al and Servant, Francisco},
  booktitle={Proceedings of the 2019 27th ACM Joint Meeting on European Software Engineering Conference and Symposium on the Foundations of Software Engineering},
  pages={397--408},
  year={2019}
}

@inproceedings{shen2021smartcommit,
  title={SmartCommit: a graph-based interactive assistant for activity-oriented commits},
  author={Shen, Bo and Zhang, Wei and K{\"a}stner, Christian and Zhao, Haiyan and Wei, Zhao and Liang, Guangtai and Jin, Zhi},
  booktitle={Proceedings of the 29th ACM Joint Meeting on European Software Engineering Conference and Symposium on the Foundations of Software Engineering},
  pages={379--390},
  year={2021}
}

@article{imtiaz2023your,
  title={Are your dependencies code reviewed?: Measuring code review coverage in dependency updates},
  author={Imtiaz, Nasif and Williams, Laurie},
  journal={IEEE Transactions on Software Engineering},
  volume={49},
  number={11},
  pages={4932--4945},
  year={2023},
  publisher={IEEE}
}

@misc{kascsite,
  title = {Understanding Vulnerability Detail Pages},
  key = {Vulnerability Detail Pages},
  howpublished = {https://nvd.nist.gov/vuln/vulnerability-detail-pages},
  year = {2025}
}

@inproceedings{wang2021patchdb,
  title={Patchdb: A large-scale security patch dataset},
  author={Wang, Xinda and Wang, Shu and Feng, Pengbin and Sun, Kun and Jajodia, Sushil},
  booktitle={2021 51st Annual IEEE/IFIP International Conference on Dependable Systems and Networks (DSN)},
  pages={149--160},
  year={2021},
  organization={IEEE}
}

@inproceedings{rosa2021evaluating,
  title={Evaluating SZZ implementations through a developer-informed oracle},
  author={Rosa, Giovanni and Pascarella, Luca and Scalabrino, Simone and Tufano, Rosalia and Bavota, Gabriele and Lanza, Michele and Oliveto, Rocco},
  booktitle={2021 IEEE/ACM 43rd International Conference on Software Engineering (ICSE)},
  pages={436--447},
  year={2021},
  organization={IEEE}
}

@inproceedings{wen2019exploring,
  title={Exploring and exploiting the correlations between bug-inducing and bug-fixing commits},
  author={Wen, Ming and Wu, Rongxin and Liu, Yepang and Tian, Yongqiang and Xie, Xuan and Cheung, Shing-Chi and Su, Zhendong},
  booktitle={Proceedings of the 2019 27th ACM Joint Meeting on European Software Engineering Conference and Symposium on the Foundations of Software Engineering},
  pages={326--337},
  year={2019}
}

@article{lyu2024evaluating,
  title={Evaluating SZZ implementations: An empirical study on the linux kernel},
  author={Lyu, Yunbo and Kang, Hong Jin and Widyasari, Ratnadira and Lawall, Julia and Lo, David},
  journal={IEEE Transactions on Software Engineering},
  volume={50},
  number={9},
  pages={2219--2239},
  year={2024},
  publisher={IEEE}
}

@article{davies2014comparing,
  title={Comparing text-based and dependence-based approaches for determining the origins of bugs},
  author={Davies, Steven and Roper, Marc and Wood, Murray},
  journal={Journal of Software: Evolution and Process},
  volume={26},
  number={1},
  pages={107--139},
  year={2014},
  publisher={Wiley Online Library}
}

@inproceedings{perl2015vccfinder,
  title={Vccfinder: Finding potential vulnerabilities in open-source projects to assist code audits},
  author={Perl, Henning and Dechand, Sergej and Smith, Matthew and Arp, Daniel and Yamaguchi, Fabian and Rieck, Konrad and Fahl, Sascha and Acar, Yasemin},
  booktitle={Proceedings of the 22nd ACM SIGSAC conference on computer and communications security},
  pages={426--437},
  year={2015}
}

@inproceedings{yang2017vuldigger,
  title={Vuldigger: A just-in-time and cost-aware tool for digging vulnerability-contributing changes},
  author={Yang, Limin and Li, Xiangxue and Yu, Yu},
  booktitle={GLOBECOM 2017-2017 IEEE Global Communications Conference},
  pages={1--7},
  year={2017},
  organization={IEEE}
}

@article{rodriguez2020bugs,
  title={How bugs are born: a model to identify how bugs are introduced in software components},
  author={Rodr{\'\i}guez-P{\'e}rez, Gema and Robles, Gregorio and Serebrenik, Alexander and Zaidman, Andy and Germ{\'a}n, Daniel M and Gonzalez-Barahona, Jesus M},
  journal={Empirical Software Engineering},
  volume={25},
  number={2},
  pages={1294--1340},
  year={2020},
  publisher={Springer}
}

@inproceedings{woo2021v0finder,
  title={$\{$V0Finder$\}$: Discovering the correct origin of publicly reported software vulnerabilities},
  author={Woo, Seunghoon and Lee, Dongwook and Park, Sunghan and Lee, Heejo and Dietrich, Sven},
  booktitle={30th USENIX Security Symposium (USENIX Security 21)},
  pages={3041--3058},
  year={2021}
}

@inproceedings{wu2024vision,
  title={Vision: Identifying affected library versions for open source software vulnerabilities},
  author={Wu, Susheng and Wang, Ruisi and Huang, Kaifeng and Cao, Yiheng and Song, Wenyan and Zhou, Zhuotong and Huang, Yiheng and Chen, Bihuan and Peng, Xin},
  booktitle={Proceedings of the 39th IEEE/ACM International Conference on Automated Software Engineering},
  pages={1447--1459},
  year={2024}
}

@inproceedings{kim2017vuddy,
  title={Vuddy: A scalable approach for vulnerable code clone discovery},
  author={Kim, Seulbae and Woo, Seunghoon and Lee, Heejo and Oh, Hakjoo},
  booktitle={2017 IEEE symposium on security and privacy (SP)},
  pages={595--614},
  year={2017},
  organization={IEEE}
}

@inproceedings{dai2021facilitating,
  title={Facilitating vulnerability assessment through poc migration},
  author={Dai, Jiarun and Zhang, Yuan and Xu, Hailong and Lyu, Haiming and Wu, Zicheng and Xing, Xinyu and Yang, Min},
  booktitle={Proceedings of the 2021 ACM SIGSAC Conference on Computer and Communications Security},
  pages={3300--3317},
  year={2021}
}

@inproceedings{dong2019towards,
  title={Towards the detection of inconsistencies in public security vulnerability reports},
  author={Dong, Ying and Guo, Wenbo and Chen, Yueqi and Xing, Xinyu and Zhang, Yuqing and Wang, Gang},
  booktitle={28th USENIX security symposium (USENIX Security 19)},
  pages={869--885},
  year={2019}
}

@article{nguyen2025toward,
  title={Toward Realistic Evaluations of Just-In-Time Vulnerability Prediction},
  author={Nguyen, Duong and Le-Cong, Thanh and Le, Triet Huynh Minh and Babar, M Ali and Huynh, Quyet-Thang},
  journal={arXiv preprint arXiv:2507.10729},
  year={2025}
}

@inproceedings{camilo2015bugs,
  title={Do bugs foreshadow vulnerabilities? a study of the chromium project},
  author={Camilo, Felivel and Meneely, Andrew and Nagappan, Meiyappan},
  booktitle={2015 IEEE/ACM 12th Working Conference on Mining Software Repositories},
  pages={269--279},
  year={2015},
  organization={IEEE}
}

@article{kamei2012large,
  title={A large-scale empirical study of just-in-time quality assurance},
  author={Kamei, Yasutaka and Shihab, Emad and Adams, Bram and Hassan, Ahmed E and Mockus, Audris and Sinha, Anand and Ubayashi, Naoyasu},
  journal={IEEE Transactions on Software Engineering},
  volume={39},
  number={6},
  pages={757--773},
  year={2012},
  publisher={IEEE}
}

@article{nguyen2024code,
  title={Code-centric learning-based just-in-time vulnerability detection},
  author={Nguyen, Son and Nguyen, Thu-Trang and Vu, Thanh Trong and Do, Thanh-Dat and Ngo, Kien-Tuan and Vo, Hieu Dinh},
  journal={Journal of Systems and Software},
  volume={214},
  pages={112014},
  year={2024},
  publisher={Elsevier}
}

@article{nguyen2016automatic,
  title={An automatic method for assessing the versions affected by a vulnerability},
  author={Nguyen, Viet Hung and Dashevskyi, Stanislav and Massacci, Fabio},
  journal={Empirical Software Engineering},
  volume={21},
  number={6},
  pages={2268--2297},
  year={2016},
  publisher={Springer}
}

@inproceedings{hanif2022vulberta,
  title={Vulberta: Simplified source code pre-training for vulnerability detection},
  author={Hanif, Hazim and Maffeis, Sergio},
  booktitle={2022 International joint conference on neural networks (IJCNN)},
  pages={1--8},
  year={2022},
  organization={IEEE}
}

@article{tang2025llm4szz,
  title={LLM4SZZ: Enhancing SZZ Algorithm with Context-Enhanced Assessment on Large Language Models},
  author={Tang, Lingxiao and Liu, Jiakun and Liu, Zhongxin and Yang, Xiaohu and Bao, Lingfeng},
  journal={Proceedings of the ACM on Software Engineering},
  volume={2},
  number={ISSTA},
  pages={343--365},
  year={2025},
  publisher={ACM New York, NY, USA}
}

@article{tang2024enhancing,
  title={Enhancing Bug-Inducing Commit Identification: A Fine-Grained Semantic Analysis Approach},
  author={Tang, Lingxiao and Ni, Chao and Huang, Qiao and Bao, Lingfeng},
  journal={IEEE Transactions on Software Engineering},
  year={2024},
  publisher={IEEE}
}

@inproceedings{bao2022vszz,
  title={V-SZZ: automatic identification of version ranges affected by CVE vulnerabilities},
  author={Bao, Lingfeng and Xia, Xin and Hassan, Ahmed E and Yang, Xiaohu},
  booktitle={Proceedings of the 44th International Conference on Software Engineering},
  pages={2352--2364},
  year={2022}
}

@article{li2021sysevr,
  title={Sysevr: A framework for using deep learning to detect software vulnerabilities},
  author={Li, Zhen and Zou, Deqing and Xu, Shouhuai and Jin, Hai and Zhu, Yawei and Chen, Zhaoxuan},
  journal={IEEE Transactions on Dependable and Secure Computing},
  volume={19},
  number={4},
  pages={2244--2258},
  year={2021},
  publisher={IEEE}
}

@inproceedings{russell2018automated,
  title={Automated vulnerability detection in source code using deep representation learning},
  author={Russell, Rebecca and Kim, Louis and Hamilton, Lei and Lazovich, Tomo and Harer, Jacob and Ozdemir, Onur and Ellingwood, Paul and McConley, Marc},
  booktitle={2018 17th IEEE international conference on machine learning and applications (ICMLA)},
  pages={757--762},
  year={2018},
  organization={IEEE}
}

@article{li2018vuldeepecker,
  title={Vuldeepecker: A deep learning-based system for vulnerability detection},
  author={Li, Zhen and Zou, Deqing and Xu, Shouhuai and Ou, Xinyu and Jin, Hai and Wang, Sujuan and Deng, Zhijun and Zhong, Yuyi},
  journal={arXiv preprint arXiv:1801.01681},
  year={2018}
}

@article{zhou2019devign,
  title={Devign: Effective vulnerability identification by learning comprehensive program semantics via graph neural networks},
  author={Zhou, Yaqin and Liu, Shangqing and Siow, Jingkai and Du, Xiaoning and Liu, Yang},
  journal={Advances in neural information processing systems},
  volume={32},
  year={2019}
}

@inproceedings{fu2022linevul,
  title={Linevul: A transformer-based line-level vulnerability prediction},
  author={Fu, Michael and Tantithamthavorn, Chakkrit},
  booktitle={Proceedings of the 19th International Conference on Mining Software Repositories},
  pages={608--620},
  year={2022}
}

@article{feng2020codebert,
  title={Codebert: A pre-trained model for programming and natural languages},
  author={Feng, Zhangyin and Guo, Daya and Tang, Duyu and Duan, Nan and Feng, Xiaocheng and Gong, Ming and Shou, Linjun and Qin, Bing and Liu, Ting and Jiang, Daxin and others},
  journal={arXiv preprint arXiv:2002.08155},
  year={2020}
}

@inproceedings{liu2024pre,
  title={Pre-training by Predicting Program Dependencies for Vulnerability Analysis Tasks},
  author={Liu, Zhongxin and Tang, Zhijie and Zhang, Junwei and Xia, Xin and Yang, Xiaohu},
  booktitle={Proceedings of the IEEE/ACM 46th International Conference on Software Engineering},
  pages={1--13},
  year={2024}
}

@misc{SARD,
      title={Software Assurance Reference Dataset},
      author={NIST},
    howpublished = {\url{https://samate.nist.gov/SARD/}}    
}

@article{chakraborty2021deep,
  title={Deep learning based vulnerability detection: Are we there yet?},
  author={Chakraborty, Saikat and Krishna, Rahul and Ding, Yangruibo and Ray, Baishakhi},
  journal={IEEE Transactions on Software Engineering},
  volume={48},
  number={9},
  pages={3280--3296},
  year={2021},
  publisher={IEEE}
}

@misc{Cppcheck,
  title = {Cppcheck - A tool for static C/C++ code analysis},
  key = {Cppcheck},
  howpublished = {https://cppcheck.sourceforge.io/},
  year = {2024}
}

@misc{Splint,
  title = {Splint Home Page},
  key = {Splint},
  howpublished = {https://splint.org/},
  year = {2024}
}

@misc{Flawfinder,
  title = {Flawfinder Home Page},
  key = {Flawfinder},
  howpublished = {https://dwheeler.com/flawfinder/},
  year = {2024}
}

@article{sate,
  title={Report on the static analysis tool exposition (sate) iv},
  author={Okun, Vadim and Delaitre, Aurelien and Black, Paul E and others},
  journal={NIST Special Publication},
  volume={500},
  pages={297},
  year={2013}
}

@article{guo2020graphcodebert,
  title={Graphcodebert: Pre-training code representations with data flow},
  author={Guo, Daya and Ren, Shuo and Lu, Shuai and Feng, Zhangyin and Tang, Duyu and Liu, Shujie and Zhou, Long and Duan, Nan and Svyatkovskiy, Alexey and Fu, Shengyu and others},
  journal={arXiv preprint arXiv:2009.08366},
  year={2020}
}

@article{guo2022unixcoder,
  title={Unixcoder: Unified cross-modal pre-training for code representation},
  author={Guo, Daya and Lu, Shuai and Duan, Nan and Wang, Yanlin and Zhou, Ming and Yin, Jian},
  journal={arXiv preprint arXiv:2203.03850},
  year={2022}
}

@inproceedings{nong2023vulgen,
  title={Vulgen: Realistic vulnerability generation via pattern mining and deep learning},
  author={Nong, Yu and Ou, Yuzhe and Pradel, Michael and Chen, Feng and Cai, Haipeng},
  booktitle={2023 IEEE/ACM 45th International Conference on Software Engineering (ICSE)},
  pages={2527--2539},
  year={2023},
  organization={IEEE}
}

@inproceedings{nong2024vgx,
  title={VGX: Large-Scale Sample Generation for Boosting Learning-Based Software Vulnerability Analyses},
  author={Nong, Yu and Fang, Richard and Yi, Guangbei and Zhao, Kunsong and Luo, Xiapu and Chen, Feng and Cai, Haipeng},
  booktitle={Proceedings of the IEEE/ACM 46th International Conference on Software Engineering},
  pages={1--13},
  year={2024}
}

@inproceedings{fan2020ac,
  title={A C/C++ code vulnerability dataset with code changes and CVE summaries},
  author={Fan, Jiahao and Li, Yi and Wang, Shaohua and Nguyen, Tien N},
  booktitle={Proceedings of the 17th International Conference on Mining Software Repositories},
  pages={508--512},
  year={2020}
}

@article{ding2024vulnerability,
  title={Vulnerability detection with code language models: How far are we?},
  author={Ding, Yangruibo and Fu, Yanjun and Ibrahim, Omniyyah and Sitawarin, Chawin and Chen, Xinyun and Alomair, Basel and Wagner, David and Ray, Baishakhi and Chen, Yizheng},
  journal={arXiv preprint arXiv:2403.18624},
  year={2024}
}

@misc{treesitter,
  title = {tree-sitter},
  key = {treesitter},
  howpublished = {https://tree-sitter.github.io/tree-sitter/index.html},
  year = {2024}
}

@article{sliwerski2005changes,
  title={When do changes induce fixes?},
  author={{\'S}liwerski, Jacek and Zimmermann, Thomas and Zeller, Andreas},
  journal={ACM sigsoft software engineering notes},
  volume={30},
  number={4},
  pages={1--5},
  year={2005},
  publisher={ACM New York, NY, USA}
}

@inproceedings{kim2006automatic,
  title={Automatic identification of bug-introducing changes},
  author={Kim, Sunghun and Zimmermann, Thomas and Pan, Kai and James Jr, E and others},
  booktitle={21st IEEE/ACM international conference on automated software engineering (ASE'06)},
  pages={81--90},
  year={2006},
  organization={IEEE}
}

@article{da2016framework,
  title={A framework for evaluating the results of the szz approach for identifying bug-introducing changes},
  author={Da Costa, Daniel Alencar and McIntosh, Shane and Shang, Weiyi and Kulesza, Uir{\'a} and Coelho, Roberta and Hassan, Ahmed E},
  journal={IEEE Transactions on Software Engineering},
  volume={43},
  number={7},
  pages={641--657},
  year={2016},
  publisher={IEEE}
}

@inproceedings{neto2018impact,
  title={The impact of refactoring changes on the SZZ algorithm: An empirical study},
  author={Neto, Edmilson Campos and Da Costa, Daniel Alencar and Kulesza, Uir{\'a}},
  booktitle={2018 IEEE 25th international conference on software analysis, evolution and reengineering (SANER)},
  pages={380--390},
  year={2018},
  organization={IEEE}
}

@inproceedings{tang2023neural,
  title={Neural SZZ algorithm},
  author={Tang, Lingxiao and Bao, Lingfeng and Xia, Xin and Huang, Zhongdong},
  booktitle={2023 38th IEEE/ACM International Conference on Automated Software Engineering (ASE)},
  pages={1024--1035},
  year={2023},
  organization={IEEE}
}

@inproceedings{wang2023graphspd,
  title={GraphSPD: Graph-based security patch detection with enriched code semantics},
  author={Wang, Shu and Wang, Xinda and Sun, Kun and Jajodia, Sushil and Wang, Haining and Li, Qi},
  booktitle={2023 IEEE Symposium on Security and Privacy (SP)},
  pages={2409--2426},
  year={2023},
  organization={IEEE}
}

@article{lu2021codexglue,
  title={Codexglue: A machine learning benchmark dataset for code understanding and generation},
  author={Lu, Shuai and Guo, Daya and Ren, Shuo and Huang, Junjie and Svyatkovskiy, Alexey and Blanco, Ambrosio and Clement, Colin and Drain, Dawn and Jiang, Daxin and Tang, Duyu and others},
  journal={arXiv preprint arXiv:2102.04664},
  year={2021}
}

@inproceedings{durumeric2014matter,
  title={The matter of heartbleed},
  author={Durumeric, Zakir and Li, Frank and Kasten, James and Amann, Johanna and Beekman, Jethro and Payer, Mathias and Weaver, Nicolas and Adrian, David and Paxson, Vern and Bailey, Michael and others},
  booktitle={Proceedings of the 2014 conference on internet measurement conference},
  pages={475--488},
  year={2014}
}

@article{luszcz2018apache,
  title={Apache struts 2: how technical and development gaps caused the equifax breach},
  author={Luszcz, Jeff},
  journal={Network Security},
  volume={2018},
  number={1},
  pages={5--8},
  year={2018},
  publisher={Elsevier}
}

@article{hiesgen2022race,
  title={The race to the vulnerable: Measuring the log4j shell incident},
  author={Hiesgen, Raphael and Nawrocki, Marcin and Schmidt, Thomas C and W{\"a}hlisch, Matthias},
  journal={arXiv preprint arXiv:2205.02544},
  year={2022}
}

@inproceedings{wu2022vulcnn,
  title={Vulcnn: An image-inspired scalable vulnerability detection system},
  author={Wu, Yueming and Zou, Deqing and Dou, Shihan and Yang, Wei and Xu, Duo and Jin, Hai},
  booktitle={Proceedings of the 44th International Conference on Software Engineering},
  pages={2365--2376},
  year={2022}
}

@inproceedings{chen2023diversevul,
  title={Diversevul: A new vulnerable source code dataset for deep learning based vulnerability detection},
  author={Chen, Yizheng and Ding, Zhoujie and Alowain, Lamya and Chen, Xinyun and Wagner, David},
  booktitle={Proceedings of the 26th International Symposium on Research in Attacks, Intrusions and Defenses},
  pages={654--668},
  year={2023}
}

@inproceedings{yamaguchi2014modeling,
  title={Modeling and discovering vulnerabilities with code property graphs},
  author={Yamaguchi, Fabian and Golde, Nico and Arp, Daniel and Rieck, Konrad},
  booktitle={2014 IEEE symposium on security and privacy},
  pages={590--604},
  year={2014},
  organization={IEEE}
}

@inproceedings{hin2022linevd,
  title={LineVD: statement-level vulnerability detection using graph neural networks},
  author={Hin, David and Kan, Andrey and Chen, Huaming and Babar, M Ali},
  booktitle={Proceedings of the 19th international conference on mining software repositories},
  pages={596--607},
  year={2022}
}

@article{zou2019mu,
  title={$\mu$ VulDeePecker: A Deep Learning-Based System for Multiclass Vulnerability Detection},
  author={Zou, Deqing and Wang, Sujuan and Xu, Shouhuai and Li, Zhen and Jin, Hai},
  journal={IEEE Transactions on Dependable and Secure Computing},
  volume={18},
  number={5},
  pages={2224--2236},
  year={2019},
  publisher={IEEE}
}

@inproceedings{xiao2020mvp,
  title={$\{$MVP$\}$: Detecting vulnerabilities using $\{$Patch-Enhanced$\}$ vulnerability signatures},
  author={Xiao, Yang and Chen, Bihuan and Yu, Chendong and Xu, Zhengzi and Yuan, Zimu and Li, Feng and Liu, Binghong and Liu, Yang and Huo, Wei and Zou, Wei and others},
  booktitle={29th USENIX Security Symposium (USENIX Security 20)},
  pages={1165--1182},
  year={2020}
}

@inproceedings{le2024latent,
  title={Are latent vulnerabilities hidden gems for software vulnerability prediction? an empirical study},
  author={Le, Triet Huynh Minh and Du, Xiaoning and Babar, M Ali},
  booktitle={Proceedings of the 21st International Conference on Mining Software Repositories},
  pages={716--727},
  year={2024}
}

@misc{nvdsite,
  title = {NVD - Home},
  key = {NVD - Home},
  howpublished = {https://nvd.nist.gov/},
  year = {2025}
}

@misc{c,
  title = {The State of Open Source Vulnerabilities 2021.},
  key = {White Source Software},
  howpublished = {https://www.whitesourcesoftware.com/open-source-vulnerability-management-report/},
  year = {2021}
}

@misc{git,
  title = {Git},
  key = {Git},
  howpublished = {https://git-scm.com/},
  year = {2025}
}

@article{voas1992pie,
  title={PIE: A dynamic failure-based technique},
  author={Voas, Jeffrey M.},
  journal={IEEE Transactions on software Engineering},
  volume={18},
  number={8},
  pages={717},
  year={1992},
  publisher={IEEE Computer Society}
}

@inproceedings{nikitopoulos2021crossvul,
  title={CrossVul: a cross-language vulnerability dataset with commit data},
  author={Nikitopoulos, Georgios and Dritsa, Konstantina and Louridas, Panos and Mitropoulos, Dimitris},
  booktitle={Proceedings of the 29th ACM Joint Meeting on European Software Engineering Conference and Symposium on the Foundations of Software Engineering},
  pages={1565--1569},
  year={2021}
}

@inproceedings{bhandari2021cvefixes,
  title={CVEfixes: automated collection of vulnerabilities and their fixes from open-source software},
  author={Bhandari, Guru and Naseer, Amara and Moonen, Leon},
  booktitle={Proceedings of the 17th International Conference on Predictive Models and Data Analytics in Software Engineering},
  pages={30--39},
  year={2021}
}

@article{fan2019impact,
  title={The impact of mislabeled changes by szz on just-in-time defect prediction},
  author={Fan, Yuanrui and Xia, Xin and Da Costa, Daniel Alencar and Lo, David and Hassan, Ahmed E and Li, Shanping},
  journal={IEEE transactions on software engineering},
  volume={47},
  number={8},
  pages={1559--1586},
  year={2019},
  publisher={IEEE}
}

@misc{5BestPractices,
  author = {{Perforce Software}},
  title = {{5 Best Git Practices for Git Commit}},
  howpublished = {{https://www.perforce.com/blog/vcs/git-best-practices-git-commit}}
}

@misc{SurveyBestPractice,
  author = {{David Golden}},
  title = {{A survey of git best practices}},
  howpublished = {{https://xdg.me/a-survey-of-git-best-practices/}}
}

@inproceedings{nguyen2013filtering,
  title={Filtering noise in mixed-purpose fixing commits to improve defect prediction and localization},
  author={Nguyen, Hoan Anh and Nguyen, Anh Tuan and Nguyen, Tien N},
  booktitle={2013 IEEE 24th international symposium on software reliability engineering (ISSRE)},
  pages={138--147},
  year={2013},
  organization={IEEE}
}

@inproceedings{sun2025dispatch,
  title={$\{$DISPATCH$\}$: Unraveling Security Patches from Entangled Code Changes},
  author={Sun, Shiyu and Xing, Yunlong and Wang, Xinda and Wang, Shu and Li, Qi and Sun, Kun},
  booktitle={34th USENIX Security Symposium (USENIX Security 25)},
  pages={4521--4540},
  year={2025}
}

@inproceedings{li2017large,
  title={A large-scale empirical study of security patches},
  author={Li, Frank and Paxson, Vern},
  booktitle={Proceedings of the 2017 ACM SIGSAC Conference on Computer and Communications Security},
  pages={2201--2215},
  year={2017}
}

@article{cheng2025vercation,
  title={VERCATION: Precise Vulnerable Open-source Software Version Identification based on Static Analysis and LLM},
  author={Cheng, Yiran and Zhang, Ting and Shar, Lwin Khin and Yang, Shouguo and Dong, Chaopeng and Lo, David and Lv, Shichao and Shi, Zhiqiang and Sun, Limin},
  journal={IEEE Transactions on Software Engineering},
  year={2025},
  publisher={IEEE}
}
